\documentclass[12pt,twocolumn]{aastex701}
\usepackage{graphics}
\usepackage{color}
\usepackage{amsmath}
\usepackage{empheq} 
\usepackage{longtable} 
\usepackage{mhchem}

\newcommand{\ltsimeq}{\la}
\newcommand{\gtsimeq}{\ga}

\newcommand{\msun}{M$_{\odot}$}
\newcommand{\mstar}{M$_*$}
\newcommand{\zsun}{Z$_{\odot}$}
\newcommand{\hi}{H{\sc i}}
\newcommand{\hii}{H{\sc ii}}

\newcommand{\mlr}{$\Upsilon_*^{3.6\mu m}$}

\newcommand{\ORF}{oxygen retention fraction}

\newcommand{\CIV}{C{\sc\ iv}}

\shortauthors{McQuinn et al.}
\shorttitle{GLOW Paper II}

\begin{document}
\title{GLOW II: A Census of Oxygen in Low-Mass Galaxies\footnote{Based on observations made with the NASA/ESA Hubble Space Telescope, obtained at the Space Telescope Science Institute, which is operated by the Association of Universities for Research in Astronomy, Inc., under NASA contract NAS 5-26555.}}

\author[0000-0001-5538-2614]{Kristen.~B.~W. McQuinn}
\affiliation{Space Telescope Science Institute, 3700 San Martin Drive, Baltimore, MD, 21218, USA}
\affiliation{Rutgers University, Department of Physics and Astronomy, 136 Frelinghuysen Road, Piscataway, NJ 08854, USA} 
\email[show]{kmcquinn@stsci.edu}

\author[0000-0003-4122-7749]{O.\ Grace Telford}
\affiliation{Department of Physics and Astronomy, University of Utah, 270 S 1400 E, Salt Lake City, UT 84112, USA}
\email{grace.telford@utah.edu}

\author[0000-0003-4307-634X]{Andrey V.~Kravtsov}
\affiliation{Dept. of Astronomy \& Astrophysics, Kavli Institute for Cosmological Physics, The University of Chicago, Chicago, IL 60637, USA}
\email{}

\author[0000-0002-1979-2197]{Joseph Burchett}
\affiliation{Department of Astronomy, 1320 Frenger Mall, New Mexico State University, Las Cruces, NM 88003, USA}
\email{}

\author[0000-0002-2970-7435]{Roger Cohen}
\affiliation{Rutgers University, Department of Physics and Astronomy, 136 Frelinghuysen Road, Piscataway, NJ 08854, USA} 
\email{}

\author[0000-0001-5368-3632]{Laura Hunter}
\affiliation{Department of Physics and Astronomy, Dartmouth College, 6127 Wilder Laboratory, Hanover, NH 03755, USA}
\email{}

\author[0000-0003-0605-8732]{Evan D. Skillman}
\affiliation{University of Minnesota, Minnesota Institute for Astrophysics, School of Physics and Astronomy, 116 Church Street, S.E., Minneapolis, MN 55455, USA}
\email{skill001@umn.edu}

\author[0000-0001-6326-7069]{Julia Roman-Duval}
\affiliation{Space Telescope Science Institute, 3700 San Martin Drive, Baltimore, MD, 21218, USA}
\email{}

\author[0000-0002-4153-053X]{Danielle A. Berg}
\affiliation{Department of Astronomy, The University of Texas at Austin, 2515 Speedway, Stop C1400, Austin, TX 78712, USA}
\affiliation{Cosmic Frontier Center, The University of Texas at Austin, Austin, TX 78712, USA} 
\email{}

\author[0000-0002-1821-7019]{John M. Cannon}
\affiliation{Department of Physics and Astronomy, Macalester College, Saint Paul, MN 55105, USA}
\email{}

\author[0000-0002-0302-2577]{John Chisholm}
\affiliation{Department of Astronomy, The University of Texas at Austin, 2515 Speedway, Stop C1400, Austin, TX 78712, USA}
\affiliation{Cosmic Frontier Center, The University of Texas at Austin, Austin, TX 78712, USA}
\email{}

\author[0000-0002-1264-2006]{Julianne J.\ Dalcanton}
\affiliation{Center for Computational Astrophysics, Flatiron Institute, 162 Fifth Avenue, New York, NY 10010, USA}
\affiliation{Department of Astronomy, University of Washington, Box 351580, Seattle, WA 98195, USA}
\email{jdalcanton@flatironinstitute.org}

\author[0000-0001-8416-4093]{Andrew E.~Dolphin}
\affiliation{Raytheon Company, 1151 E. Hermans Road, Tucson, AZ 85756, USA}
\affiliation{University of Arizona, Steward Observatory, 933 North Cherry Avenue, Tucson, AZ 85721, USA}
\email{}

\author{Liese van Zee}
\affiliation{Department of Astronomy, Indiana University, 727 East Third Street, Bloomington, IN 47405, USA}
\email{}

\author[0000-0002-7502-0597]{Benjamin Williams}
\affiliation{Department of Astronomy, University of Washington, Box 351580, Seattle, WA 98195, USA}
\email{}

\begin{abstract}
Oxygen is forged by stars and redistributed through galaxies by feedback-driven outflows, leaving a record of star formation and the baryon cycle imprinted on its present-day abundance and distribution. The Galaxies Losing Oxygen via Winds (GLOW) project quantifies the production, distribution, and retention of oxygen in 37 low-mass, low-metallicity, gas-rich galaxies in the nearby universe ($D<6$ Mpc) spanning a critical stellar mass range ($10^{6.5}<$\mstar$<10^{9.5}$) where feedback is expected to strongly shape galaxy evolution. We incorporate literature results to extend the analysis over $10^{5}<$\mstar$<10^{11.5}$. We couple star formation histories derived from resolved stars with measurements of ISM metallicity and gas mass, and compares the oxygen budgets with empirical constraints in the circumgalactic medium (CGM). We find low-mass galaxies retain between 4$-$25\% (mean 11\%) of their oxygen produced by stellar nucleosynthesis, and nearly all retained oxygen residing in the gas. Galaxies exhibit increased scatter to higher oxygen retention fractions when residing in denser environments ($\Theta_5>0.5$) and within 1 Mpc of a massive galaxy (\mstar$>10^9$ \msun), indicating environment likely affects the amount of oxygen retained, recycled, or accreted to galaxies. Contrary to expectations, oxygen retention does not correlate with position on the mass-metallicity relation. Although ionized oxygen is detected in the CGM, it remains unclear whether most missing oxygen resides there or has been expelled entirely. Hydrodynamical simulations, despite successfully reproducing the mass-metallicity relation, predict much higher retention fractions than measured in low-mass galaxies. Simple modeling indicates that wind mass-loading and outflow metallicity govern oxygen retention, and that low-mass galaxies accrete less baryonic material relative to the cosmic baryon fraction.
\end{abstract} 


\section{Introduction}\label{sec:intro}
Galaxy evolution is fundamentally driven by the process of turning gas into stars and the subsequent return of chemically enriched material and energy. Metals produced by star formation are distributed back into the interstellar medium (ISM), and, given sufficient energy from stellar feedback, can be driven out of a galaxy into the circumgalactic medium (CGM) and beyond via outflows. The cumulative impact of this process is thought to be mass dependent. Lower mass galaxies have lower star formation rates and efficiencies and produce correspondingly fewer metals than more massive spirals, but they are also thought to lose a greater percentage of their metals via outflows because of their shallower potential \citep[e.g.,][]{Larson1974, Matteucci1983, Dekel1986, Recchi2001, Dalcanton2007, Brooks2007, Finlator2008, Peeples2011, Muratov2015, McQuinn2015b, Chisholm2018, Christensen2018, Maiolino2019}.

Indeed, at lower galaxy masses, stellar-feedback driven outflows are thought to be a major factor in determining a galaxy's present-day metallicity \citep{Dalcanton2007}. Yet the loss of metals via outflows, and the final destination of those metals, remains poorly understood. Outflow properties, including their velocities, densities, chemical composition, and physical trajectories, can vary significantly; thus, metals ejected in an outflow can cycle back to the host galaxy on relatively short timescales, build-up as a reservoir in a galaxy's CGM, or be lost completely to the intergalactic medium (IGM). 

Adding complexity to understanding the fate of the metals, galaxies also interact with their immediate surroundings at the CGM/IGM interface (via ram pressure stripping but also gas accretion), and through tidal forces from neighboring galaxies \citep{Burchett:2016aa,Hasan:2024_slimeWeb}. Thus, the local environment of a galaxy may also play a role in determining where metals end up, how metals cycle through galaxies, and possibly whether metals are transferred between neighbors \citep{Hafen:2020_fateCGM}.

The combined effects of astration and outflows, the complexities related to local environment, and to a lesser extent gas inflows \citep[e.g.,][]{Edmunds1990, Garnett2002, Dalcanton2007}, is thought to set the mass-metallicity (MZ) scaling relation, where there is a tight correlation between gas-phase oxygen abundances and stellar masses in galaxies \citep{Tremonti2004, Lee2006, Berg2012, Andrews2013, Zahid2014,  Mannucci2010, Curti2020, Sanders2021}. However, while the MZ relation captures the bulk properties of a population, scaling relations do not fully isolate the phenomena from which the population properties emerge, nor do they explain what drives galaxies away from the mean to set the dispersion of the relation. The dispersion about the MZ relation is the largest in the low-mass galaxy regime \citep[e.g.,][]{McQuinn2020, Breneman2025}, making dwarfs galaxies the best laboratories for probing the drivers of the MZ relation and learning about the cycling of material through galaxies. 

One way to uniquely and quantitatively trace the baryon cycle --- which informs the physics driving the MZ relation --- is to track the distribution of metals in individual galaxies \citep{Kirby2011, McQuinn2015b, Telford2019}. Metals are forged by stars, returned to the ISM, and removed from the host galaxy primarily by supernovae \citep[e.g.,][]{Larson1974, Dekel1986, Tremonti2004}. A census of the metals in a galaxy therefore unambiguously records the net sum of star formation and supernova feedback in low-mass galaxies\footnote{Note that AGN feedback may become important in driving metal-loss from galaxies for the most massive (e.g., \mstar\ $>10^9$ \msun) dwarf galaxies \citep{Dashyan2018}.}.

Here, the Galaxies Losing Oxygen via Winds (GLOW) project aims to quantify the baryon cycle by measuring the production, distribution, and retention of metals in a statistically significant sample of gas-rich, star-forming dwarf galaxies spanning three decades in stellar mass (\mstar\ $\sim10^{6.5} - 10^{9.5}$ \msun). We focus on oxygen as a tracer of the metals in the galaxies. Oxygen has a relatively simple nucleosynthetic pathway that is well-understood \citep[e.g.,][]{Timmes1995}, is the most abundant heavy element, can be readily measured in the gas-phase with optical spectra \citep[e.g.,][]{Skillman1989, Lee2006, Berg2012}, and is considered a direct tracer of material expelled in supernova ejecta. 

In this work, we build on empirical results from McQuinn et al.\ (2026; hereafter Paper I) to produce a comprehensive census of oxygen in low-mass galaxies. We combine detailed empirical constraints on the distribution of oxygen within galaxies (i.e., constraints on the oxygen content in the gas, dust, stellar disks, and the CGM) with measurements of the various galaxy properties (e.g., global \hi, star, metallicity values, gravitational potential), to provide unique insights into the baryon cycle driving galaxy evolution at the low-mass end of the galaxy mass function. 

The GLOW galaxies are located at distances between $\sim1-6$ Mpc, which places them within the Local Sheet of galaxies \citep{Tully2008}. Furthermore, they are dispersed both inside and outside of the known ring of galaxies, which is an over-density of brighter, more massive galaxies at $3-5$ Mpc \citep{McCall2014}. Within in the Local Sheet environment, the GLOW galaxies are broadly considered `field' galaxies in that they are not satellites of massive galaxies. Yet, the density of the local environments of the GLOW sample vary and have a range in tidal field strengths (measured by the tidal index parameter; see \S\ref{sec:env}), similar to differences within the Local Sheet itself \citep{Neuzil2020}. For example, some of the GLOW galaxies are located in galaxy groups (e.g., the M~81 and Canes Venetici I Group) and others are truly isolated (e.g., $>$2 Mpc from any known galaxy with \mstar\ $>10^9$ \msun). Thus, we also investigate how the metal content of low-mass galaxies may vary as a function of environment, albeit over a relatively narrow range of environment densities.

In this paper, we use the data products presented in Paper I from the GLOW project to trace the production and distribution of oxygen in 37 low-mass galaxies in the nearby (D $<6$ Mpc) universe. In Section~\ref{sec:budget}, we summarize the overall data sets and usages, and then present the detailed calculations for determining the oxygen produced and contained in the gas, the stars and stellar remnants, and the dust. In Section~\ref{sec:results}, we present the percent of oxygen retained as a function of various galaxy quantities (\mstar, rotational velocity from the \hi\ data, gas-to-star ratio, and gas-phase oxygen abundance), explore the retention fraction with the MZ relation, and the impact the local environment plays in determining the retention of oxygen in galaxies. In Section~\ref{sec:discuss}, we then discuss the possible metal content of the CGM, connect our results to those in the literature to explore the oxygen budget across the stellar mass range of \mstar\ $= 10^5 - 10^{11.5}$ \msun, compare these results to hydrodynamical cosmological simulations, and explore different outflow scenarios using a simple regulator-type model of galaxy evolution of \citet{Kravtsov.Manwadkar.2022}. Finally, in Section~\ref{sec:conclusions}, we summarize our main takeaways.

\section{The Oxygen Budget Calculation}\label{sec:budget}
In this section, we begin with an overview of the datasets used in the analysis (\S~\ref{sec:overview_data}). We then describe the calculations to determine the total mass of oxygen produced by stellar nucleosynthesis (\S~\ref{sec:Oproduced}) and the mass of oxygen contained in the gas (\S~\ref{sec:Ogas}), stars and stellar remnants (\S~\ref{sec:Ostars}), and dust (\S~\ref{sec:Odust}). Finally, we describe the calculus for estimating the percent of oxygen retained in the disks of the galaxies (\S~\ref{sec:mrf}).
\subsection{Overview of Datasets}\label{sec:overview_data}

Measuring the total oxygen produced by stellar nucleosynthesis and estimating the current oxygen content in different galaxy components requires data from myriad observatories. As mentioned above, these data were reduced and analyzed in Paper ~I and supplemented with measurements reported in the literature. To help guide the reader, we provide a summary of the datasets presented in Paper~I and how they are used in the present work. Detailed descriptions of individual calculations and related assumptions are provided throughout the rest of the Section~\ref{sec:budget}.

Briefly, the production of oxygen is based on the stellar nucleosynthetic yield of oxygen from core-collapse supernovae (CCSNe) and the total stellar mass estimated from 3.6~$\micron$ imaging assuming a mass-to-light ratio. The oxygen in the gas is estimated using the direct-method on optical spectral of the \hii\ regions and the gas mass contained with 4.4 scale lengths of the stellar component based on galaxy geometry determined from 3.6~$\micron$ imaging. The oxygen in the stars and stellar remnants combines the mass of stars formed as a function of time with the age-metallicity relations derived from fitting color-magnitude diagrams (CMDs) of resolved stars and assuming an alpha to iron ratio. The oxygen in the dust is estimated from the expected depletion of oxygen as function of metallicity and the neutral hydrogen column density. Constraints on the metals in the CGM come from UV absorption line spectroscopic measurements (using background quasars as a light source) of low-mass galaxies obtained from the literature. 

Supplementing the data needed for the oxygen budget calculations, we collate measurements from the literature on galaxy properties (including, for example, \hi\ properties, accurate distances based on the tip of the red giant branch method, nitrogen abundances, etc.) which are used in our analysis. Finally, using data from the Updated Nearby Galaxy Catalog \citep{Karachentsev2013}, we quantify the local environment around each galaxy. For convenience, in Table~\ref{tab:Properties} we summarize the galaxy properties derived from the from data presented in Paper I.

\begin{deluxetable*}{llccccccc}
\tabletypesize{\footnotesize}
\label{tab:Properties}
\tablecaption{Summary of Galaxy Properties}
\tablehead{
\colhead{Galaxy} & \colhead{Distance} & \colhead{log} & \colhead{log} & \colhead{$f_{gas}$} & \colhead{12+log(O/H)} & \colhead{V$_{\rm rot}$} & \colhead{$\Theta_5$} & \colhead{NLG Sep} \\
\colhead{}	& \colhead{(Mpc)}	&\colhead{(\mstar/\msun)}	&  \colhead{(M$_{\rm HI}$/\msun)}	& \colhead{(M$_{\rm gas}$/(M$_{gas} +$ \mstar))}	& \colhead{} & \colhead{km s$^{-1}$} & \colhead{}	& \colhead{(Mpc)} 
} 
\colnumbers
\startdata
WLM      & 0.98$^{+0.04}_{-0.02}$ & 7.68 $\pm$ 0.12 & 7.84 & 0.66 & 7.77 $\pm$ 0.10 & 32.00 & -1.04 & 0.89  \\
UGC00685 & 4.81$^{+0.04}_{-0.04}$ & 7.89 $\pm$ 0.12 & 7.84 & 0.54 & 8.00 $\pm$ 0.03 & 43.00 & \nodata & 2.35  \\
NGC0625  & 4.02$^{+0.07}_{-0.06}$ & 8.77 $\pm$ 0.12 & 8.07 & 0.21 & 8.08 $\pm$ 0.12 & 37.00 & -0.87 & 1.43 \\
NGC0784  & 5.37$^{+0.02}_{-0.05}$ & 8.65 $\pm$ 0.12 & 8.59 & 0.54 & 7.97 $\pm$ 0.05 & \nodata  & -1.32 & 2.43  \\
NGC1569  & 3.19$^{+0.09}_{-0.09}$ & 8.98 $\pm$ 0.12 & 8.31 & 0.22 & 8.15 $\pm$ 0.05 & 33.00 & 1.18  & 0.31 \\
NGC2366  & 3.28$^{+0.03}_{-0.05}$ & 8.40 $\pm$ 0.12 & 8.65 & 0.70 & 7.91 $\pm$ 0.05 & \nodata  & 1.09  & 0.21  \\
UGC04305 & 3.47$^{+0.03}_{-0.03}$ & 8.46 $\pm$ 0.12 & 8.63 & 0.66 & 7.92 $\pm$ 0.10 & 38.00 & 1.02  & 0.55  \\
UGC04459 & 3.68$^{+0.03}_{-0.05}$ & 7.31 $\pm$ 0.12 & 7.81 & 0.81 & 7.82 $\pm$ 0.09 & \nodata & 0.99  & 0.53  \\
UGC04483 & 3.58$^{+0.15}_{-0.15}$ & 6.97 $\pm$ 0.12 & 7.40 & 0.78 & 7.56 $\pm$ 0.03 & 19.00 & 1.48  & 0.11  \\
UGC05139 & 4.02$^{+0.06}_{-0.04}$ & 7.84 $\pm$ 0.12 & 8.30 & 0.79 & 7.92 $\pm$ 0.05 & \nodata & 1.62  & 0.32  \\
UGC05364 & 0.74$^{+0.05}_{-0.16}$ & 6.42 $\pm$ 0.12 & 6.73 & 0.73 & 7.30 $\pm$ 0.05 & 9.00 & -1.93  & 1.51 \\
SextansB & 1.43$^{+0.02}_{-0.02}$ & 7.57 $\pm$ 0.12 & 7.59 & 0.58 & 7.84 $\pm$ 0.05 & 25.00 & -1.51 & 2.14 \\
NGC3109  & 1.34$^{+0.05}_{-0.06}$ & 8.14 $\pm$ 0.12 & 8.38 & 0.70 & 7.77 $\pm$ 0.07 & \nodata & -0.27  & 2.00  \\
SextansA & 1.45$^{+0.05}_{-0.05}$ & 7.52 $\pm$ 0.12 & 7.80 & 0.72 & 7.54 $\pm$ 0.09 & 41.00 & -1.12 & 2.22 \\
UGC05666 & 3.93$^{+0.04}_{-0.05}$ & 8.84 $\pm$ 0.12 & 9.09 & 0.70 & 7.93 $\pm$ 0.05 & \nodata & 1.90   & 0.26 1 \\
NGC3738  & 5.30$^{+0.05}_{-0.05}$ & 8.74 $\pm$ 0.12 & 8.13 & 0.24 & 8.04 $\pm$ 0.06 & 45.00 & -2.74 & 1.03 \\
NGC3741  & 3.22$^{+0.18}_{-0.16}$ & 7.15 $\pm$ 0.12 & 7.34 & 0.67 & 7.68 $\pm$ 0.03 & 58.00 & -1.42 & 1.52 \\
UGC06817 & 2.65$^{+0.10}_{-0.10}$ & 7.32 $\pm$ 0.12 & 7.76 & 0.79 & 7.53 $\pm$ 0.02 & 16.00 & -0.69 & 1.72  \\
NGC4068  & 4.39$^{+0.04}_{-0.04}$ & 8.05 $\pm$ 0.12 & 8.23 & 0.67 & 8.00 $\pm$ 0.10 & 30.00 & -0.67 & 0.75  \\
NGC4163  & 2.99$^{+0.03}_{-0.04}$ & 7.64 $\pm$ 0.12 & 7.25 & 0.35 & 7.56 $\pm$ 0.14 & 16.00 & 1.62  & 1.94 \\
CGCG269- & 4.61$^{+0.19}_{-0.17}$ & 7.00 $\pm$ 0.12 & 7.44 & 0.79 & 7.47 $\pm$ 0.02 & 16.00 & -2.83 & 1.78  \\
UGCA281  & 5.70$^{+0.13}_{-0.11}$ & 7.57 $\pm$ 0.12 & 7.71 & 0.65 & 7.80 $\pm$ 0.03 & 32.00 & -0.16 & 0.69  \\
UGC07577 & 2.61$^{+0.06}_{-0.05}$ & 7.82 $\pm$ 0.12 & 7.54 & 0.41 & 7.97 $\pm$ 0.06 & 15.00 & -0.92 & 1.67 \\
NGC4449  & 4.27$^{+0.02}_{-0.02}$ & 9.40 $\pm$ 0.12 & 9.07 & 0.38 & 8.32 $\pm$ 0.03 & 63.00 & 0.51  & 0.54  \\
UGCA292  & 3.85$^{+0.09}_{-0.55}$ & 7.16 $\pm$ 0.12 & 7.67 & 0.81 & 7.30 $\pm$ 0.03 & 23.00 & -0.86 & 0.93 \\
UGC08024 & 4.04$^{+0.06}_{-0.07}$ & 7.47 $\pm$ 0.12 & 8.12 & 0.86 & 7.67 $\pm$ 0.06 & 49.00 & 0.43  & 0.51  \\
UGC08091 & 2.19$^{+0.12}_{-0.09}$ & 6.73 $\pm$ 0.12 & 6.95 & 0.69 & 7.65 $\pm$ 0.06 & 16.00 & -3.39 & 2.10 \\
UGC08201 & 4.83$^{+0.04}_{-0.04}$ & 8.01 $\pm$ 0.12 & 8.19 & 0.67 & 7.80 $\pm$ 0.06 & 24.00 & -0.45 & 0.56  \\
UGC08508 & 2.67$^{+0.10}_{-0.10}$ & 7.28 $\pm$ 0.12 & 7.29 & 0.58 & 7.76 $\pm$ 0.07 & 28.00 & -2.03 & 1.51 \\
UGC08638 & 4.29$^{+0.04}_{-0.06}$ & 7.59 $\pm$ 0.12 & 7.30 & 0.40 & 7.94 $\pm$ 0.05 & 17.00 & -0.13 & 0.77 \\
UGC08651 & 3.10$^{+0.06}_{-0.06}$ & 7.25 $\pm$ 0.12 & 7.47 & 0.69 & 7.85 $\pm$ 0.04 & 22.00 & -1.15 & 1.64  \\
NGC5253  & 3.44$^{+0.02}_{-0.03}$ & 8.94 $\pm$ 0.12 & 8.05 & 0.15 & 8.15 $\pm$ 0.10 & 33.00 & 0.52  & 0.75 \\
UGC09128 & 2.30$^{+0.04}_{-0.03}$ & 6.85 $\pm$ 0.12 & 7.19 & 0.74 & 7.75 $\pm$ 0.05 & 20.00 & -2.77 & 2.23  \\
UGC09240 & 2.83$^{+0.04}_{-0.05}$ & 7.59 $\pm$ 0.12 & 7.58 & 0.56 & 7.95 $\pm$ 0.03 & 34.00 & -2.10 & 1.69  \\
IC4662   & 2.55$^{+0.02}_{-0.02}$ & 8.18 $\pm$ 0.12 & 7.50 & 0.22 & 8.17 $\pm$ 0.04 & 51.00 & \nodata  & 1.92 \\
NGC6789  & 3.55$^{+0.07}_{-0.05}$ & 7.65 $\pm$ 0.12 & 7.17 & 0.30 & 7.83 $\pm$ 0.05 & 52.00 & \nodata & 2.36 \\
IC5152   & 1.96$^{+0.05}_{-0.03}$ & 8.25 $\pm$ 0.12 & 7.70 & 0.27 & 7.92 $\pm$ 0.07 & 47.00 & -1.32 & 0.89  \\
\enddata
\tablecomments{Galaxies in the sample and summary of properties reported in McQuinn et al.\ (2026), reproduced here for convenience. M$_{\rm HI}$ is the \hi\ mass contained within 4.4 scale lengths. $f_{gas}$ is based on the \hi\ and stellar masses in Cols. 2 and 3, with a 1.33 correction for helium in the gas.}
\end{deluxetable*}

\subsection{Oxygen Produced by Stellar Nucleosynthesis}\label{sec:Oproduced}
The total mass of oxygen produced by stars is estimated using the total stellar mass (\mstar) ever formed and assuming a constant yield of newly synthesized oxygen per solar mass of gas converted to stars ($P$). The total stellar mass formed is based on  3.6~$\micron$ fluxes measured from surface brightness profiles fit to contaminant-cleaned IRAC images and extrapolating the profiles to infinity. We then adopt a mass-to-light ratio, \mlr, of 0.47 \citep{McGaugh2014} derived for a Salpeter initial mass function (IMF), apply a factor of 0.85 to convert to a Kroupa IMF mass scale, assume a solar magnitude at 3.6~$\micron$ (M$_{3.6;\rm solar}$) of 3.24 mag \citep{Oh2008}, and adopt the TRGB-based distance moduli ($\mu_{0}$) to the galaxies. We assume an uncertainty of 0.12 dex on the log of the stellar masses. This uncertainty is based on the range in mass-to-infrared light ratios at 3.6 \micron\ of 0.45$-$0.6 (i.e., a
range of 0.15, which translates to an uncertainty of 33\% in stellar mass or 0.12 dex); it also corresponds to the upper end of the uncertainties quoted for various mass-to-light ratio calibrations at 3.6 \micron\ \citep[e.g.,][]{Zhu2010, McGaugh2014, Meidt2014, Schombert2019}. The present-day stellar mass calculation is shown in Equation~\ref{eq:mstar}:
\begin{equation}
M_* = \Upsilon_*^{3.6\mu m}  \times 0.85 \times (10^{(m_{3.6;\rm total} - M_{3.6;\rm solar} - \mu_{0})/2.5}) .
\label{eq:mstar}
\end{equation}

We then adopt an oxygen yield (P) value of 0.010 from \citet{Nomoto2013}, which is calculated for a low gas-phase metallicity ($Z=0.004$, cf.\ mean value of 0.002 for the GLOW sample) in a stellar population averaged over a Kroupa IMF in the mass range of 10 to 40 \msun. Finally, to account for the material that has been returned to the ISM via stellar winds and supernovae (i.e., to convert from a present-day stellar mass to a measure of the total stellar mass formed), we scale the stellar mass values up by (1-R), where R is the return mass fraction defined as the total mass fraction returned to the ISM by a stellar generation and accounts for mass locked in stellar remnants \citep{Vincenzo2016}; see Paper~I, Section~4 for additional details. In sum, the oxygen mass produced by stellar nucleosynthesis is calculated using Equation~\ref{eq:oproduce}:
\begin{equation}
M^O_{\rm produced} = {\rm P} \times M_* / (1 - R). 
\label{eq:oproduce}
\end{equation}

\noindent We use a return mass fraction R of 0.43 based on modeling stellar population synthesis at low metallicities, which also assumes a Kroupa IMF \citep{Vincenzo2016}, consistent with the assumption used for both oxygen yield and the SFH derivations, and that preserves the mass in stellar remnants. As noted in \citet{Vincenzo2016}, the fraction of gas returned can vary slightly as a function of metallicity and age of the stellar populations. However, for the range of properties of the GLOW sample, this has less than a 5\% impact on the derived masses, which is well within our mass uncertainties and does not impact our results. Thus, the adoption of a constant value is justified.

\subsection{Oxygen in the Gas}\label{sec:Ogas}
The oxygen mass in the gas is calculated based on the abundance of oxygen measured from the temperature-sensitive [\ion{O}{3}] $\lambda4363/\lambda5007$ line ratio (i.e., the ``direct method''), and the measured atomic gas masses in the galaxies. This approach assumes that the oxygen is well-mixed in the galaxy and that the  oxygen abundance measured from individual \hii\ regions is representative of the ISM. Although ISM enrichment from supernovae can be a local phenomenon, spatial variations in the oxygen abundance are rarely observed as the mixing volumes are large. Indeed, while there are exceptions \citep{James2020}, the majority of observations show that metallicity variations and gradients are small to non-existent in dwarf galaxies \citep{Skillman1989, Kobulnicky1996, Lee2006, Croxall2009, Berg2012}. Further evidence supporting well-mixed ISMs across the disks comes from a study of \hii\ regions from $r<r_{25}$ to $r\sim 2r_{25}$ in lower-mass SINGS galaxies where the slopes of the oxygen abundances as a function of radius were consistent with zero \citep{Werk2011}. 

However, the assumption of a well-mixed ISM breaks down when a gaseous disk is significantly more extended than the stellar component. For example, in the GLOW galaxy NGC3741, the \hi\ extends more than 20$\times$ the scale length of the stars measured from 3.6~$\micron$ imaging (Paper~I) and the gas in the outer regions of this system is likely to have experienced significantly less chemical enrichment compared with the gas in the inner regions co-located with the stars \citep[e.g.,][]{Garnett2002}. Therefore, to avoid over-estimating the amount of oxygen in the gas, we exclude atomic gas at large radii relative to the stars. We adopt a characteristic stellar extent of 4.4 $\times$ the scale-length ($\alpha$) and include the mass of \hi\ within this region based on the galaxy geometry, as calculated in Paper I. For convenience, we briefly summarize the process here. We use 4.4$\alpha$ measured from the 3.6~$\micron$ surface brightness profiles to set the extent of the gaseous disk to consider as enriched. While 3.3$\alpha$ is typically referenced as enclosing the higher surface brightness regions of galaxies, the stellar component is clearly observed out to 4.4$\alpha$ in all cases, which encloses $\sim$93.4\% of the total light assuming an exponential profile. Our full calculation to determine the oxygen mass in the gas then becomes:
\begin{equation}
M^O_{\rm gas} = (O/H) \times (m_O/m_p) \times M_{\rm HI; 4.4\alpha}.
\label{eq:ogas}
\end{equation}

\noindent where (O/H) is based on the gas-phase oxygen abundance (12+log(O/H) values), $m_{\rm O}/m_{\rm p}$ represent the atomic mass ratio of oxygen to the proton, and $M_{\rm HI; 4.4\alpha}$ is the mass of atomic hydrogen within 4.4 scale lengths based on the 3.6~$\micron$ imaging. The majority of the galaxies have most of their \hi\ within this radius (see Figure~10; Paper I); thus, setting this limit has little impact on the calculated retention fractions in these systems. For the few galaxies with extended \hi\ disks, the retention fractions could be higher than we calculate if appreciable amounts of oxygen have been recycled to the galaxies and reside at larger radii. We experimented with including the total gas mass in the calculations for the galaxies with extended \hi\ disks, which would assume an unlikely constant metallicity throughout the outer disk, and found that the retention fractions were greater than 100\% in all cases, which is unphysical. Thus, we proceed with using the limit of 4.4$\alpha$, which is a more reasonable assumption describing the fraction of \hi\ that experiences significant enrichment, and note higher uncertainties for the galaxies with extended gaseous disks.

\subsection{Oxygen Locked in Stars and Stellar Remnants}\label{sec:Ostars}
The oxygen mass in the stars and stellar remnants is estimated using the metallicity evolution of the stars (the age-metallicity relation, Z(t)) and the star formation histories (SFR(t)). Z(t) and SFR(t) are derived simultaneously by fitting the CMDs of the resolved stars from Hubble Space Telescope optical imaging with stellar evolution libraries and assuming a number of parameters (e.g., IMF, binary fraction, distance, etc.; see Paper~I for details). The fits for SFR(t) and Z(t) were performed using two different stellar evolution libraries, namely PARSEC \citep{Bressan2012} and MIST \citep{Dotter2016, Choi2016}; the differences in solutions provide an estimate of the systematic uncertainties. Here, we use the results from the PARSEC models as a basis for our primary analysis, but include results from MIST in Appendix~\ref{sec:appendix}. The approach of using metallicity estimates from the CMDs is an alternative to determining the metallicities of the stars spectroscopically, which is not feasible given the distances to the galaxies and current observing facilities.

We use the Z(t) values derived from the CMDs, which represents an average [M/H] per time bin informed by stellar populations of different ages. We note that inferring [M/H] from photometry has inherently higher uncertainties than spectroscopically determined abundances due to degeneracies in the CMD. However, CMD-based Z(t) results have been shown to be a reliable measure of the metallicity evolution. For example, in a recent study of the Local Group galaxy WLM, the metallicity distribution function (MDF) resulting from the SFR(t) and Z(t) of red giant branch stars was in excellent agreement with the MDF based on spectroscopically measured [Fe/H] from 44 red giant branch stars in a similar region of the CMD \citep{McQuinn2024, Leaman2009}. Recent studies of the LMC and SMC also show good agreement between spectroscopic abundance constraints and the CMD-based Z(t) \citep{Cohen2024a, Cohen2024b}. Further confirmation is provided by a comparison of the average present-day [M/H] from the CMD-fitting approach with the spectroscopically measured gas-phase oxygen abundances in the system (Paper I, Figure 9). The overall agreement between methods and the uncertainties on individual metallicities also supports binning the CMD-based metallacities to an average.  

To convert from [M/H] to [O/H], we assume that [M/H] is representative of [Fe/H] and adopt a [O/Fe] $= 0$ representative of the solar ratio. While this is a reasonable assumption, stars with lower [Fe/H] metallicity values have been noted to have an enhanced amount of oxygen and, more generally, enhanced amount of all $\alpha$ elements. The trend of higher [$\alpha$/Fe] ratios at lower [Fe/H] ratios is mainly driven by changes in a galaxy's SFH and the relative number of Type II supernovae from young massive stars (the primary source of $\alpha$ elements with some contribution of Fe) to Type Ia supernovae from older populations \citep[which contributes significant amounts of Fe, see, e.g.,][]{Gilmore1991}. Detailed chemical abundance studies in star-forming galaxies over a range of stellar masses show $\alpha$-enhancements of $0.2-0.3$ dex in low metallicity ([Fe/H] $\sim -1.5$) stars (e.g., Leo~A, Aquarius, SagDIG) and $0.1-0.2$ enhancements at slightly more metal-rich ([Fe/H] $\sim -1.0$) environments (e.g., the Large and Small Magellanic Clouds) \citep{Tolstoy2009, Kirby2017, Hasselquist2017, Skuladottir2017, Hill2019, Nidever2020, Hasselquist2021}. 

Therefore, to account for potential variation in [O/Fe] ratio, we propagate an additional 0.2 dex uncertainty in the value of Z(t). Note that since the majority of the oxygen resides in the ISM, the uncertainties from this conservative assumption, and the uncertainties in our approach of using Z(t) based on CMD-fitting, do not limit our ability to discern between different models in our final oxygen budget calculations.

Our calculation for the mass of oxygen locked in stars and stellar remnants multiplies the stellar mass formed in each time bin with the corresponding stellar metallicity value and assumes a return mass fraction ($R$) of gas from the stellar populations, as shown in Equation~\ref{eq:O_stars}:

\begin{equation}
M^O_{*} = (1 - R)\,\sum ~SFR(t)~Z(t)\,~\delta t , \label{eq:O_stars}
\end{equation}

\noindent where Z(t) is measured from fitting the CMDs and is converted to oxygen mass density based on the following:
\begin{equation}
\log (Z(t)) = \log(O(t)/H) + \log\frac{m_{\rm H} \times 16}{m_{\rm H}\times X + m_{\rm He} \times Y}. \end{equation}

\noindent and assuming $X$, the fraction of hydrogen, is 75\% and $Y$, the fraction of helium, is 25\%. We use the same return fraction R of 0.433 applied above in the stellar mass calculations (see Equation~\ref{eq:oproduce}). 

As noted above, the oxygen mass produced by stars is based on the total stellar mass estimated from the 3.6~$\micron$ fluxes measured from surface brightness profiles extrapolated to infinity and assuming a mass-to-light ratio. However, the oxygen retained in the stars and stellar remnants is based on measurements from the HST optical imaging of the resolved stellar populations contained within the HST footprints. In many cases, the areal coverage of the HST footprint encompasses the extended stellar disk of the galaxies, but there are several exceptions. 

To ensure we are accounting for any stellar component outside the HST footprints, we take a straightforward approach and use the total stellar mass from the 3.6$\micron$ fluxes to determine the mass `missed' by the HST observations. We then calculate the oxygen locked in the stars in these outer regions by multiplying the additional mass in the 3.6$\micron$ images with the earliest value of Z(t) (i.e., assuming the outer regions of the galaxies will be populated by older stars). Summing the oxygen in the stellar component measured from these two different methods is justified as stellar masses measured from matched apertures based on the 3.6$\micron$ fluxes and CMD-fitting from the stellar populations were shown to be in very good agreement with each other (Paper I). In addition, while our approach ensures completeness in the calculation, it does not appreciably change the oxygen content estimated in the stars given the smaller fraction of stars at larger radii and the metal-poor nature of the extended stellar populations.

\subsection{Oxygen in the Dust}\label{sec:Odust}
Even in the low-metallicity galaxies that comprise our sample, oxygen in the ISM can deplete onto dust grains and must be accounted for in the oxygen budget calculation. We estimate the oxygen mass in the dust using the empirical framework developed by \citet{Jenkins2009} for relating the depletion of various elements in the ISM (in particular iron and oxygen), and the calibrations of iron depletion as a function of metallicity and gas density based on the Magellanic Clouds \citep{RomanDuval2022} and on lower metallicity galaxies \citep{Hamanowicz2024}. These works showed that for a fixed metallicity, dust growth is more efficient in regions where the column density of \hi\ is higher (i.e., the gas is denser), and, thus, there is greater depletion of an element. In lower metallicity gas, the depletion for a fixed gas density decreases as dust growth becomes less efficient with fewer metals available for grain growth. These previous studies also showed that oxygen is only mildly depleted compared to refractory elements such as Fe, Si, and Mg.

Here, we summarize how we apply the above framework to calculate the mass of oxygen in dust, and refer the reader to the original works for more details on the approach and the empirical calibrations. Briefly, the depletion strength of oxygen ($\delta(O)$) can be parameterized using three empirically calibrated coefficients, $A_{\rm{O}}$, $B_{\rm{O}}$, and $z_{\rm{O}}$, and a factor $F_*$ that is determined from the neutral hydrogen column density and the metallicity of the ISM:

\begin{equation}
\delta(\rm{O}) = B_{\rm{O}} + A_{\rm{O}}(F_* - z_{\rm{O}}),
\label{eq:dust_depletion}
\end{equation}

To determine $\delta(\rm{O})$, we use the $A_{\rm{O}}$, $B_{\rm{O}}$, and $z_{\rm{O}}$ coefficients for oxygen from \citet{Jenkins2009}, the depletion as a function of \hi\ column density and metallicity from \citet{Hamanowicz2024} and \citet{RomanDuval2022}, the gas-phase oxygen abundances of the galaxies from Table~\ref{tab:Properties}, and the \hi\ column densities across the spatially resolved galaxies using interferometric observatories (see Paper~I). The \hi\ column density maps allow us to calculate iron depletions on a pixel-by-pixel basis for galaxies in different metallicity bins from Table 7 in \citet{Hamanowicz2024} and Table 3 of \citet{RomanDuval2022}. From the maps of iron depletions, we then derive maps of $F_*$ from the $A_{\rm{Fe}}$, $B_{\rm{Fe}}$, and $z_{\rm{Fe}}$ coefficients in \citet{Jenkins2009}. Equation \ref{eq:dust_depletion} is then use to convert $F_*$ to a map of oxygen depletions, $\delta$(O). Finally, we estimate the mass of oxygen in the dust, $M_{dust}^O$, by summing the oxygen depleted from the spatially resolved maps:

\begin{equation}
M_{dust}^{\rm{O}} = (1-10^{\delta(\rm{O})}) \left ( \frac{\rm{O}}{\rm{H}} \right ) m_{\rm{O}} M(HI)
\end{equation}

\noindent where $M$(\hi) is the mass of hydrogen in each pixel. Note that the depletion of oxygen drops sharply when the \hi\ column density is below a certain threshold \citep[e.g., $N_{HI} \ltsimeq$ a few $\times 10^{20}$ atoms~cm$^{-2}$; see][for details]{RomanDuval2022, Hamanowicz2024}. As a result, in a few cases where the \hi\ column densities are overall low, the amount of oxygen locked in dust is negligible and are reported as zero.

\subsection{Missing Oxygen Mass and Oxygen Retention Fraction Calculations}\label{sec:mrf}
With estimates of the oxygen produced by stars and the subsequent distribution of oxygen in the gas, dust, stars, and stellar remnants, we can now calculate the oxygen mass missing from the galaxies with Equation~\ref{eq:Odeficit} and the corresponding fraction of oxygen retained in the systems with Equation~\ref{eq:orf}:

\begin{equation}
\Delta M^O = M^O_{total} - M^O_{gas} - M^O_{dust} - M^O_* \label{eq:Odeficit}
\end{equation}

\begin{equation}
{\rm Fraction~Retained} = (M^O_{\rm gas} + M^O_{\rm dust} + M^O_*) /  M^O_{\rm total} \label{eq:orf}
\end{equation}

\noindent Values of the oxygen masses and retention fractions in the various components are provided in Table~\ref{tab:Obudget}.

\section{Empirical Results}\label{sec:results}
In this section, we present the results on the oxygen mass in the different galaxy components for the GLOW sample (\S\ref{sec:Omass}), and explore the oxygen retention fractions as a function of effective yield (\S\ref{sec:effective_yields}), galaxies properties (\S\ref{sec:compare_properties}), the MZ relation (S\ref{sec:compare_mzr}), and the local environments of the galaxies (\S\ref{sec:env}).

\subsection{Estimates of the Oxygen Mass}\label{sec:Omass}
\begin{figure*}
\includegraphics[width=0.98\textwidth]{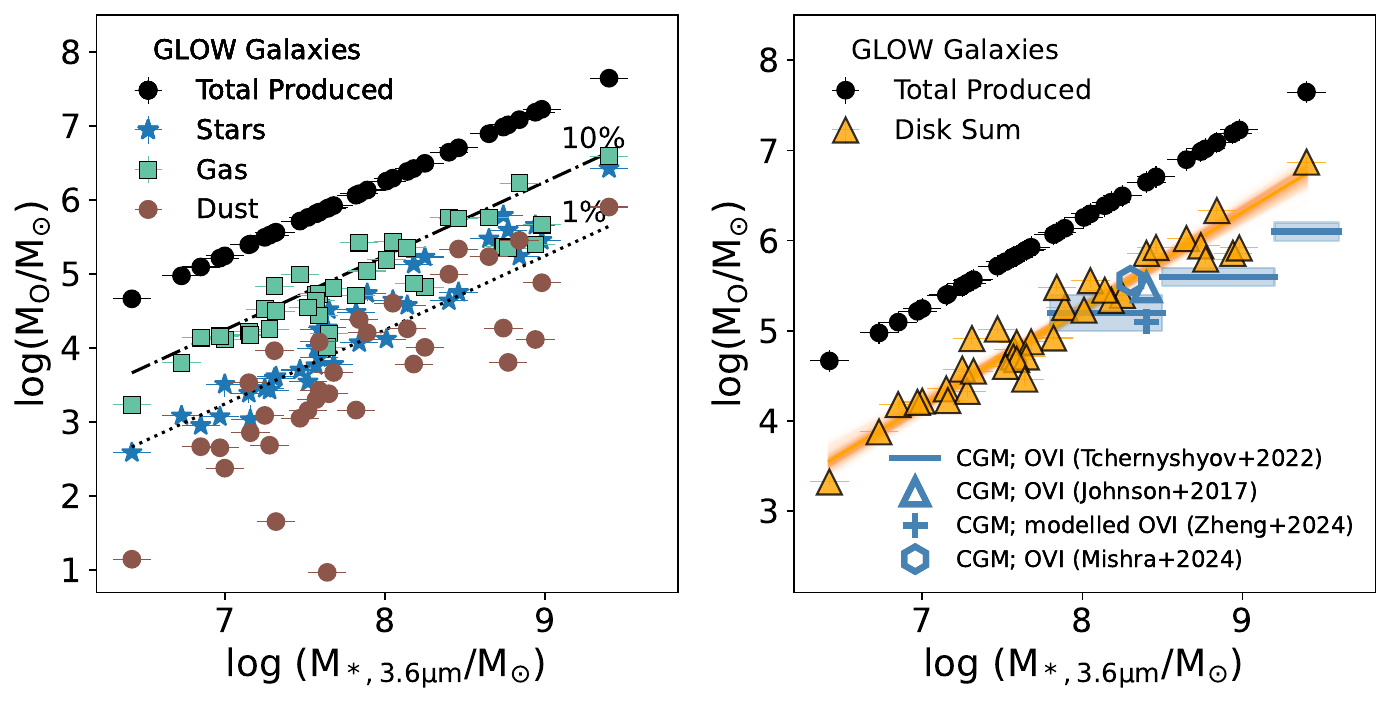}
\caption{{\em Left panel:} The logarithm of mass of oxygen, $\log(M_{\rm O}/M_*$), produced by stellar nucleosynthesis (black), and the mass of oxygen measured in the stars (blue), gas (green), and dust (brown) in each of the GLOW galaxies plotted as a function of the stellar mass of the galaxies. To guide the eye, we show black lines representing 10\% (dash-dotted) and 1\% (dotted) oxygen retention fractions. Uncertainties on stellar mass and total oxygen produced take into account the uncertain conversion in the \mlr; uncertainties on each subcomponent are described above in the individual sections on oxygen calculations. {\em Right panel:} The mass of oxygen produced (black) compared with the oxygen in the disk (orange), which is simply the sum of the oxygen mass measured in the gas, dust, and stars, as a function of stellar mass. We also overplot the best-fitting line to the total oxygen retained in the disk as a thick orange with realizations within 1$\sigma$ confidence intervals as thinner orange lines that take into account uncertainties on the mass of oxygen in each of the subcomponents. The gap between the oxygen produced and the mass present in the disks reflects the significant amount of oxygen `missing'. Also shown are representative constraints of the mass of \ion{O}{6} retained in the CGM of dwarf galaxies from four difference studies from the literature (light blue).}
\label{fig:orf_components}
\end{figure*}

Figure~\ref{fig:orf_components} (left panel) tallies the total mass of oxygen produced by star formation and the portion of that mass estimated to be in each of the galaxy subcomponents (stars, gas, dust), as a function of stellar mass of the galaxies. All values are provided in Table~\ref{tab:Obudget}. To help guide the reader, we also overlay black lines representing 10\% (dash-dotted) and 1\% (dotted) of the oxygen mass to highlight the overall low oxygen retention fractions in the galaxies and give context to the amount of oxygen present in the different subcomponents. 

The total oxygen mass produced is a perfect linear relation with stellar mass, given that it is calculated assuming a constant yield of oxygen from the stars (Equation~\ref{eq:oproduce}). However, the mass in different subcomponents varies among the galaxies. Among the various subcomponents, the majority of oxygen resides in the gas (green squares), followed by the stars (blue stars), and then, finally, the dust (brown circles). Perhaps surprisingly, while oxygen content in the stars is greater than in the dust, the stars still make up only a minor contribution to the oxygen mass budget. The gas itself (Equation~\ref{eq:ogas}) has the dominant contribution to the mass of oxygen in the galaxies and, as a consequence, this component makes the largest contribution to any random or systematic uncertainties.

Figure~\ref{fig:orf_components} (right panel) adds the subcomponents together to show the current total mass of oxygen in the galaxy disk ($M^O_{gas} + M^O_{dust} + M^O_*$), plotted as orange triangles. There is a significant gap between the oxygen mass produced (black circles, with uncertainties from the stellar mass measurements) and the oxygen mass in the disk (orange triangles; best-fitting line with realizations within
1$\sigma$ confidence). This gap reflects the significant amount of oxygen that was produced, but that is no longer present in the disk of the galaxies. 

Overplotted in Figure~\ref{fig:orf_components} (right panel) for the upper mass range of our sample are estimates of the metal content thought to be retained in the CGM of galaxies. The estimates shown are based on four separate studies that have constrained the mass of a high-ionization state of oxygen, OVI, in the CGM of dwarfs at low redshifts \citep{Johnson2017, Tchernyshyov2022, Zheng2024, Mishra2024}. We discuss the CGM measurements in detail in the context our results in Section~\ref{sec:cgm}.

\subsection{Retention Fraction and Effective Yields}\label{sec:effective_yields}
In a closed box model of galaxy evolution (i.e., no gas inflows or outflows), as gas is converted into stars, the gas fraction ($f_{gas}\equiv M_{\rm gas}/(M_{\rm gas}+$\mstar)) decreases while the metallicity ($Z$) increases. This basic principle is encapsulated in the relation $Z = y_{\rm true} \times \ln(1/f_{\rm gas})$ \citep{Searle1972}, where the yield is considered the `true' yield as it is defined as the mass in primary elements freshly produced by massive stars present in stars and stellar remnants in a galaxy and is a constant of $\sim0.02$ (i.e., for every unit of stellar mass formed, $\sim$2\% of the mass is in metals, depending on the IMF and nucleosynthetic yield adopted).

However, if pristine gas is accreted or metals are lost, the true yields decreases to the `effective' yield, $y_{\rm eff}$ defined as:

\begin{equation}
y_{\rm eff} = \frac{Z}{ \ln[1/f_{gas})]} \label{eq:yeff}
\end{equation}

\noindent which provides a measure of how much a system deviates from a closed box model, with lower values of $y_{\rm eff}$ indicating greater inflows of pristine gas and/or metal-enriched outflows. For the GLOW galaxies, in Equation~\ref{eq:yeff}, we determine $Z$ based on the gas-phase oxygen abundance assuming a solar 12$+$log(O/H) $=8.76$ or \zsun $=$ 0.01524 \citep{Bressan2012} (consistent with the values used in the PARSEC stellar libraries), and the gas fraction, $f_{gas}$ using the gas mass within 4.4 scale lengths (M$_{HI;4.4\alpha}$) scaled by 1.33 to account for helium\footnote{Given the paucity of detections of molecular gas in low-metallicity galaxies and the uncertainties associated with molecular gas mass estimates, we do not include the possible molecular content in the galaxies here.}. The resulting values of $y_{\rm eff}$ range from $0.001$ to $0.01$, all below the expected true yield value.

\begin{figure}
\includegraphics[width=0.48\textwidth]{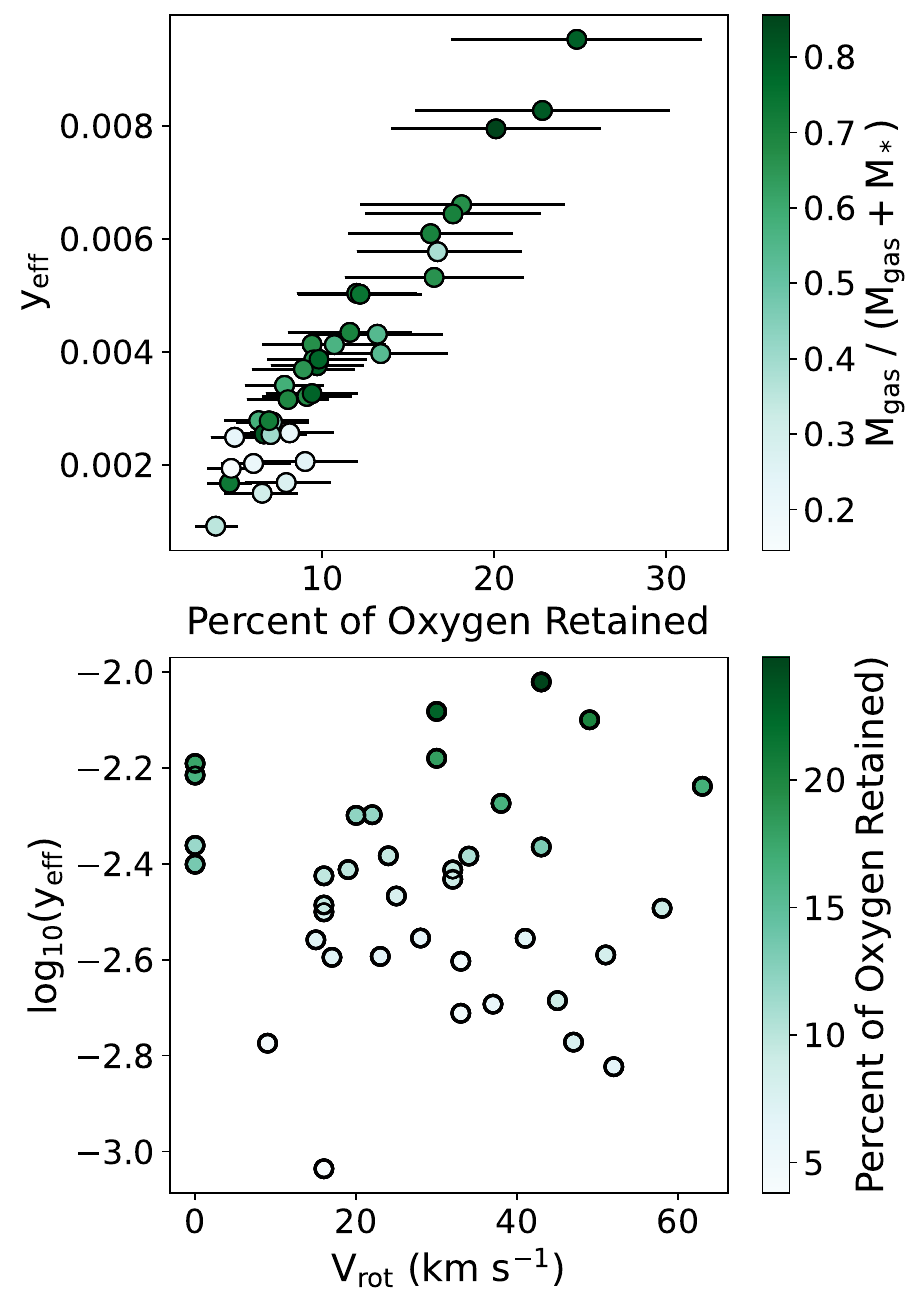}
\caption{Top Panel: The effective yields as a function of the percent of oxygen retained in the galaxies (in the stars, gas, and dust) for the GLOW sample, colored-coded by the $f_{gas}$. As expected, galaxies with lower retention fractions have lower effective yields and, generally, lower gas fractions. Bottom Panel: The logarithm of the effective yields as a function of rotational velocity, color-coded by the percent of oxygen retained in the galaxies.}
\label{fig:orf_yeff}
\end{figure}

The top panel of Figure~\ref{fig:orf_yeff} presents the effective yields ($y_{\rm eff}$) as a function of the percent of oxygen retained for the GLOW galaxies, color-coded by $f_{gas}$; the bottom panel shows the log of the effective yields (log$_{10}$(y$_{\rm eff}$)) as a function of \hi\ rotation velocity, now color-coded by the percent of oxygen retained. From the top panel, we find that galaxies with low effective yields have higher gas fractions, similar to \citet{Dalcanton2007}. We also find that the range of effective yields for the GLOW galaxies is similar at the low end, but extends to higher values ($-$2.0 here, versus $-$2.5 in \citet{Dalcanton2007}, or $-$2.7 if restricted to the mass range of the GLOW galaxies).

The relationship in this top panel is extremely tight, which can be explained as follows. In the limit where the retained oxygen mass is dominated by the gas with a negligible amount of oxygen in the stars and dust, the oxygen retained in the disk is $M^O_{disk}\approx\,Z\,M_{gas}$. Since the total oxygen production is $M_O= P\,M_*/(1-R)$ by definition, the retained fraction can be written as $f_{retain}=M^O_{disk}/M_O \approx (Z/P)\,(1-R)\,(M_{gas}/M_*)$. If the denominator in Equation~\ref{eq:yeff} for the effective yield is rewritten as $\ln{(1 + M_*/M_{gas})}$, then $y_{\rm eff}\approx Z/\ln{(1 + (1/f_{retain})\,(Z/P)\,(1-R))}$. In the limit where $M_*<M_{gas}$, a Taylor-series expansion yields a linear relationship between $y_{\rm eff}$ and $f_{retain}$, with no dependence on metallicity or gas fraction (i.e., $y_{\rm eff}\approx\,f_{retain}(P/(1-R)$). Thus, in gas-rich galaxies where the stars and dust become inconsequential reservoirs of oxygen compared to the gas, then the effective yield should have a monotonic, nearly linear relationship with the oxygen retention fraction, as indeed is seen in the top panel of Figure~\ref{fig:orf_yeff}.

The top panel of Figure~\ref{fig:orf_yeff} illustrates that the effective yields are lower for lower oxygen retention fractions, indicating that metal outflows (rather than gas inflows) are driving the low effective yields for these gas-rich galaxies. See \citet{Dalcanton2007} Section~4 for a full discussion of the impact of gas inflows vs.\ outflows in gas-rich galaxies.

From Figure~\ref{fig:orf_yeff}, the \ORF\ values are overall quite low. The mean \ORF\ for the sample is 10.7\% with a standard deviation of 5.1\%; the lowest value is $\sim3.8$\% and the highest value is close to an order of magnitude higher at 24.8\%. $\sim$80\% (29/37) of galaxies have an \ORF\ $\leq15$\% while the remaining $\sim$20\% have retained fractions of $\sim$15-25\%, although note that the largest retained fractions also have the largest uncertainties and are all statistically consistent with a value of 20\%. 

The bottom panel of Figure~\ref{fig:orf_yeff} shows no clear trend between y$_{\rm eff}$ and rotational velocity, which is in tension with the simple model that supernova feedback is more effective at driving metals out of galaxies at lower masses with shallower potential wells. Earlier work on effective yields by \citet{Garnett2002} showed decreasing y$_{\rm eff}$ with lower rotational velocities, but the sample size of low mass galaxies was small (of order a dozen), there was significant scatter at low masses, and the trend was more established given the range of velocities probed was much larger (e.g., up to $\sim$300 km s$^{-1}$). We also find no trend between the rotational velocities and gas fraction, in contrast to the results for lower mass galaxies shown in \cite{Dalcanton2007}.

\begin{figure}
\includegraphics[width=0.48\textwidth]{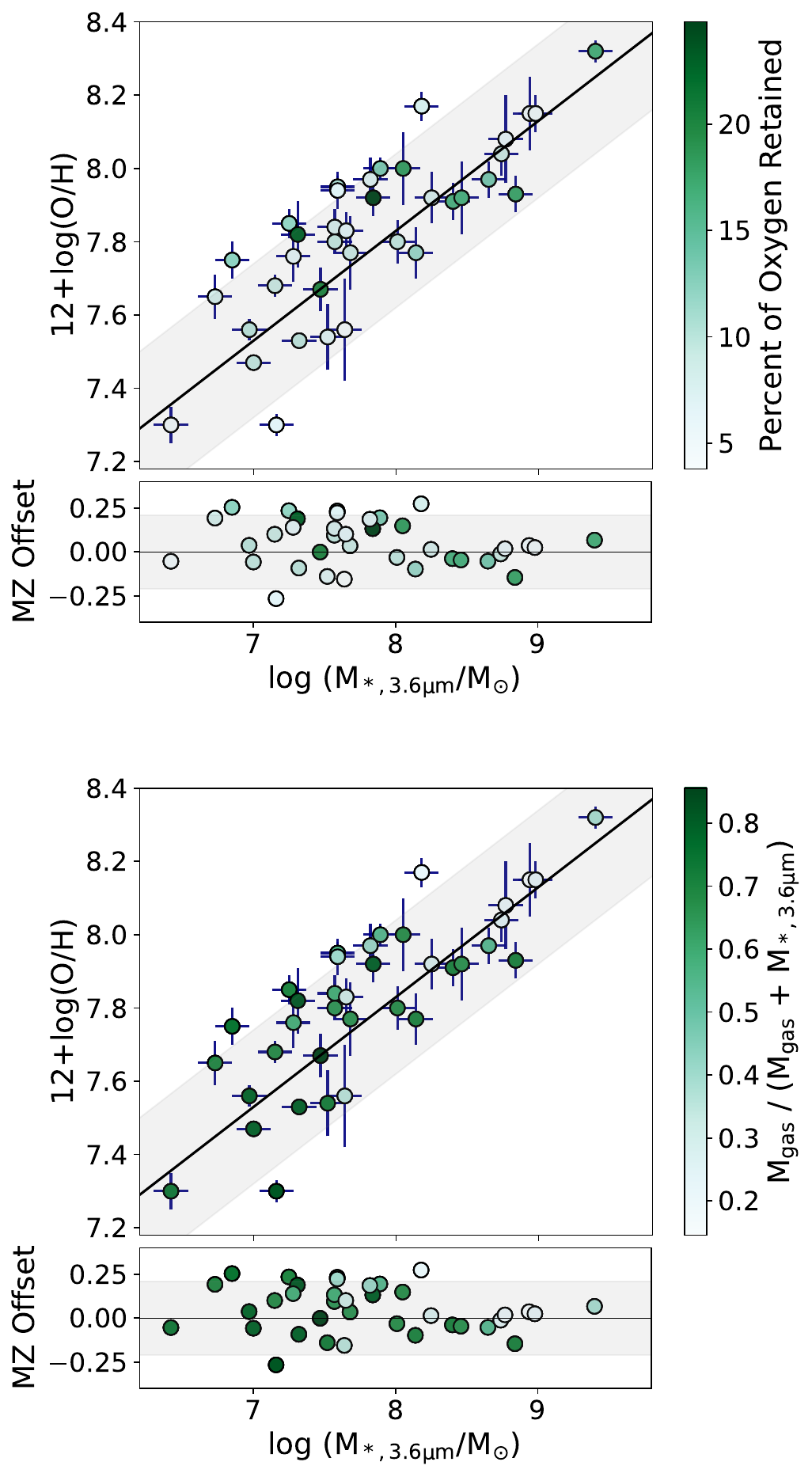}
\caption{Mass-metallicity (MZ) relation for GLOW galaxies color-coded by the percent of oxygen retained (top) and $f_{gas}$ (bottom). The black line and gray shaded regions represent the mean MZ relation and 1$\sigma$ dispersion measured from Local Volume galaxies \citep{Berg2012}. The smaller panels below each plot show the perpendicular offset from each point to that mean relation; the black horizontal line demarcates no offset and shaded regions represent the disperson on the MZ relation. While the MZ relation is thought to be driven by a combination of astration, star formation efficiency, and metal loss at low galaxy masses, there is no strong correlation between the MZ relation and the fraction of metals retained in the galaxies. There is also no clear trend of gas-phase oxygen abundance with gas-fraction, although the well-known trend of increasing gas-richness of galaxies at lower masses is recovered.}
\label{fig:orf_mzr}
\end{figure}

\subsection{Mass-Metallicity Relation and Retention Fractions}\label{sec:compare_mzr}
The trend of increasing gas-phase oxygen abundance with stellar mass, i.e., the MZ relation, is thought to arise from the combined effects of astration, star formation efficiency, and the decreasing ability of low-mass galaxies to retain their metals. Indeed, the increased efficacy of galactic winds to remove metals in dwarf galaxies is often invoked to explain the steep decline in the MZ relation for low-mass galaxies. 

Empirical constraints support this scenario. From a sample of seven galaxies, the metallicity of galactic outflows relative to the native ISM metallicity has been estimated to be $10-50\times$ higher in two low-mass galaxies ($10^{7-8}$ \msun) but only $2.6\times$ higher in 5 more massive systems ($10^{9-11}$ \msun), indicating that outflows in lower mass galaxies may more efficiently remove metals relative to higher mass systems \citep{Chisholm2018}. This same study estimated that the metal-loading factors in the two low-mass galaxies reach values close to $\sim100$, implying that the galaxies are losing 100$\times$ more metals than are retained in the recently formed stars. Similar metal-loading factors have been reported from the MAGMA survey which studied more than 300 local star-forming galaxies \citep{Tortora2022}. 

The high {\em metal}-loading factors are in contrast to the more modest {\em mass}-loading factors measured in low-mass galaxies (see Section~\ref{sec:regulatory_model}), indicating that metals synthesized by stars are preferentially ejected from low-mass galaxies whereas the bulk of the gas is retained and/or recycled via local fountains. This framing also implies different channels for metal loss: metals can be preferentially expelled in a hot-wind phase with higher concentrations than found in the overall ISM of a galaxy, while the loss of any bulk ISM mass is via a cooler wind phase and can be quite modest. The loss of metals is also likely occurring on smaller, local scales around supernovae events, whereas the loss of ISM material may occur via larger galactic wind events which are expected to have metallicities consistent with the ISM metallicity. This has been confirmed anecdotally:  the gas-phase oxygen abundance of the wind in a nearby galaxy NGC~1569,  which is also included in the GLOW sample, was measured to be consistent with the ISM metallicity \citep{HamelBravo2024}.

Figure~\ref{fig:orf_mzr} explores this framework and attempts to connect metal retentions and gas-content with the MZ relation for the GLOW galaxies by color-coding the galaxies with the percent of oxygen retained (top panel) and $f_{gas}$ (bottom panel). Given the expectation that metal loss is a driver of the MZ relation, one might expect that the scatter in metallicity for a given stellar mass might be correlated with retention fraction, but that is not the case. While there is some indication that galaxies with lower metallicities also have low retention fractions, there is no clear trend in the MZ plane (top panel). 

This is shown explicitly in the subpanel which quantifies the perpendicular offset of each galaxy from the mean MZ relation; there is no discernible trend in offset from the MZ relation. For example, for a stellar mass of $\sim10^9$ \msun, galaxies with higher gas-phase abundances also show lower retention fractions (contrary to expectations) while at a lower mass of $\sim10^7$ \msun, the gas-phase abundances span $\sim0.5$ dex without a significant difference in retention fractions (see also Figure~\ref{fig:orf_properties}, bottom panel). While the mass-dependent nature of metal expulsion is clearly an important driver in determining the metallicity of low-mass galaxies, the lack of a discernible trend and the large scatter implies there must also be other factors that help set the MZ relation. The color-coding of the MZ relation by $f_{gas}$ (bottom panel) reveals the well-known trend of increasing gas-richness at lower galaxy mass, but shows little trend between gas-richness and gas-phase oxygen abundance. 

\subsection{Retention Fraction as Function of Galaxy Properties}\label{sec:compare_properties}
\begin{figure*}
\begin{center}
\includegraphics[width=0.415\textwidth]{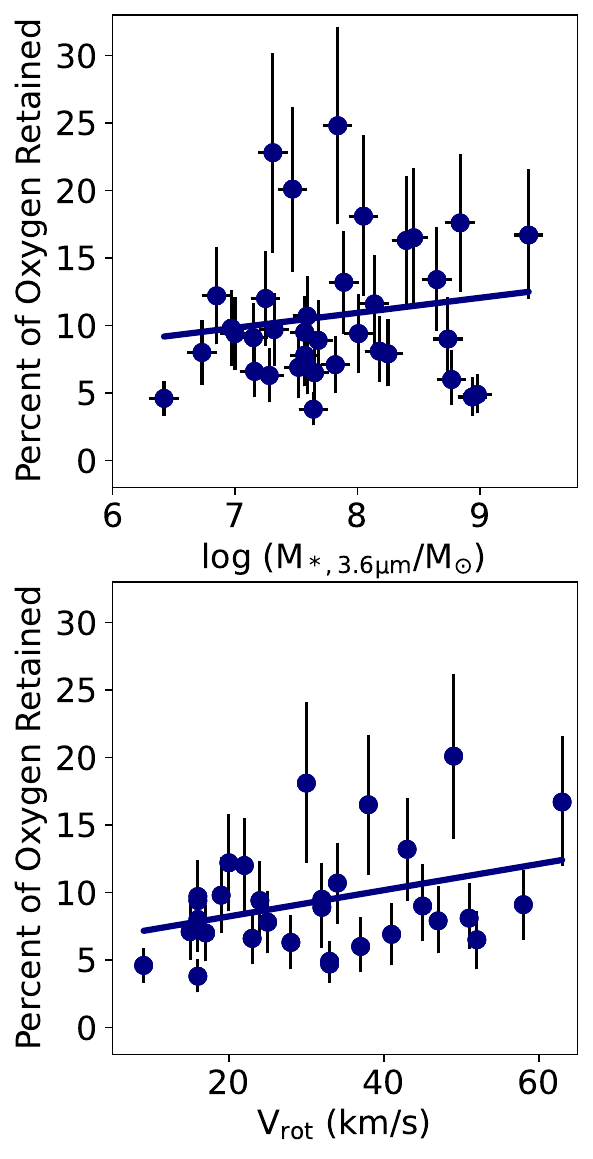}
\includegraphics[width=0.48\textwidth]{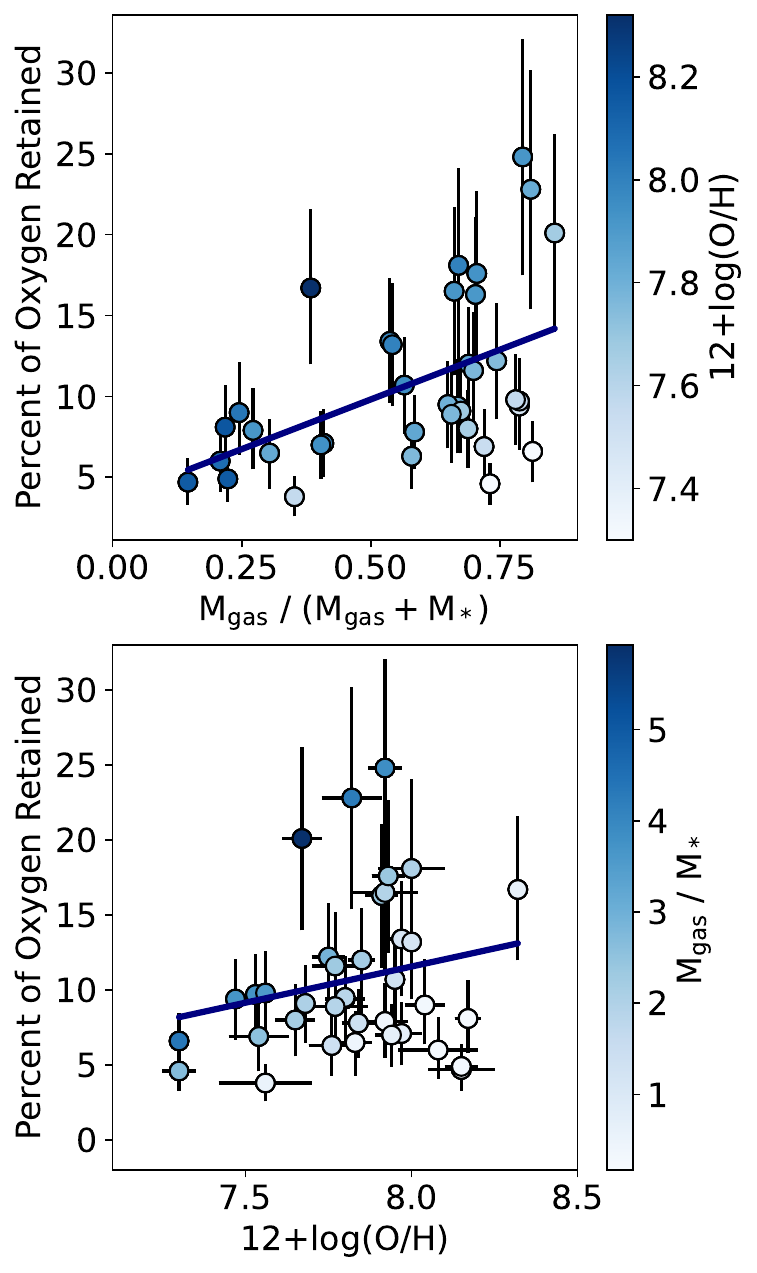}
\end{center}
\caption{Percent of oxygen retained as a function of galaxies stellar mass, \mstar\ (top left) and the rotational velocity of the \hi\ (V$_{\rm rot}$; bottom left), which is a measure of the gravitational potential, $f_{gas}$ color-code by oxygen abundance (top right), and gas-phase oxygen abundance, 12+log(O/H), color-coded by gas ratio (M$_{\rm gas}/$\mstar; bottom right). To highlight overall trends, we also show best-fitting lines in both panels.}
\label{fig:orf_properties}
\end{figure*}

Figure~\ref{fig:orf_properties} shows the percentage of oxygen retained as a function of various galaxy properties including: stellar mass, $M_*$, gas rotational velocity (V$_{\rm rot}$) based on the full width at half maximum of the 21 cm line measurements (W50) and corrected for inclination angle, $f_{gas}$, and gas-phase oxygen abundance, 12+log(O/H). In each panel, circles show the values of individual galaxies; overall trends in the data are highlighted by best-fit lines. Uncertainties on the individual points are based on propagating uncertainties on the stellar mass calculations, the SFHs, the gas-phase oxygen abundances, the \hi\ masses, and the possible range of O/Fe ratios.

In the top left panel of Figure~\ref{fig:orf_properties}, the retention fractions shows a modest decrease with decreasing stellar mass. In the bottom left panel, we recover a similar trend of decreasing retention fractions with decreasing rotational velocities (i.e., shallower gravitational potentials). While it is possible to convert the rotational velocities to dynamical masses as a proxy for halo mass, given the uncertainties in the stellar mass-halo mass relation at low masses and the limited extent of \hi\ typically measured low-mass galaxies \citep[e.g.,][]{McQuinn2022} that are also disparate within the sample, we prefer a more direct comparison of the retention fractions with the empirically determined gas rotational velocities. Note there is somewhat larger scatter seen in the trend with stellar mass to that seen with the rotational velocity, which suggests a modest amount of scatter in the stellar mass-halo mass relation in this mass regime, as expected.

In the top right panel of Figure~\ref{fig:orf_properties}, there is a clear trend of an increasing percentage of oxygen retained with increasing gas-richness; this is just another way to visualize the trend with effective yields shown in Figure~\ref{fig:orf_yeff}. The color-coding represents the gas-phase oxygen abundance is lower in galaxies with greater gas content and provides a different visualization of the parameters that set the effective yields in Equation~\ref{eq:yeff} and the simplified linear relation between effective yields and the oxygen retention fraction (see Section~\ref{sec:effective_yields}). In the bottom right panel, we find a modest trend between retention fractions and gas-phase oxygen abundances. Here, the points are color-code by gas {\em ratios}, showing that the lower retention fractions are found in galaxies with the lowest gas ratios. We also found no correlation between \ORF\ and the log(N/O) ratios for the galaxies (not shown).

Table~\ref{tab:Obudget} provides all values based on the PARSEC models. Appendix~\ref{sec:appendix} presents sibling plot to Figure~\ref{fig:orf_properties} with the \ORF\ values based on SFHs derived from the MIST stellar libraries, instead of the PARSEC libraries shown here; the differences in the calculated retention fractions from the two stellar libraries provides a consistency check on the SFH and $Z(t)$ values and gives an indication of level of the systematic uncertainties in the stellar models. The differences in overall results are minimal as the fraction of oxygen retained in the stars is such a small component relative to the oxygen produced and the oxygen retained in the gas. 

\begin{figure}
\begin{center}
\includegraphics[width=0.48\textwidth]{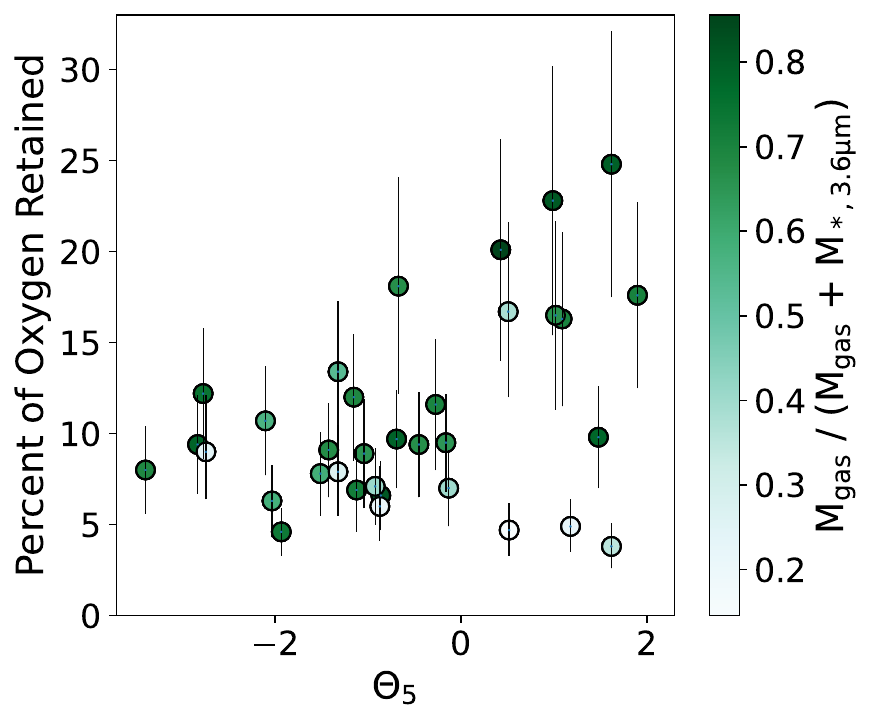}
\includegraphics[width=0.48\textwidth]{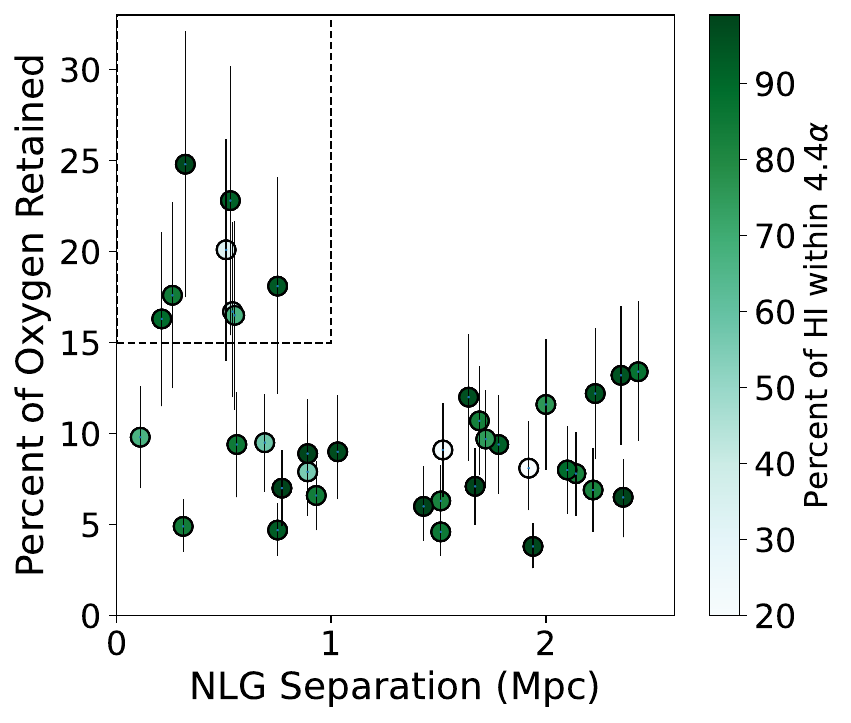}
\end{center}
\caption{Top panel: Percent of oxygen retained as a function of the  density of the local environment measured via the tidal index ($\Theta_5$), color-coded by $f_{gas}$. There is a general trend that galaxies in denser environments retain a greater fraction of their oxygen. The exceptions are the galaxies with lower $f_{gas}$ which do not retain many metals at all. Bottom panel: Percent of oxygen retained as a function of the separation from the nearest large (\mstar $>10^9$\msun) galaxy (NLG) color-coded by the fraction of the \hi\ mass measured within 4.4$\alpha$ of the stars. The most isolated galaxies consistently show low retention fractions, whereas the retention fractions for galaxies closer to a nearby massive system, which also have higher values of $\Theta_5$, show significant scatter. Despite the more complex, denser environments of these galaxies, we find no evidence of \hi\ disk truncation; this is not unexpected given that the sample was designed to only include gas-rich galaxies that are overall considered to be in the field.}
\label{fig:orf_env}
\end{figure}

\subsection{Environment and Retention Fractions}\label{sec:env}
Figure~\ref{fig:orf_env} explores the percent of oxygen retained in the galaxies and their gas content as a function of two metrics that quantify the environment around a galaxy and which were previously presented in Paper~I, Table~9. The first metric is the tidal index parameter ($\Theta_5$), which is a measure of the overall density of the surrounding environment and relates the tidal force experienced by a galaxy ($i$) from the closest $n$ galaxies, where we adopt $n$ to be five:
\begin{equation}
    \Theta_5 = \textrm{max}[\textrm{log}(M_{n} / D_{i,n}^{3})] + C  \textrm{, $n$ = 1, 2, ..., 5}
\end{equation}\label{eq:tidal_index}

\noindent In the calculation of $\Theta_5$, $M_{n}$ is the mass of the $n$th galaxy, $D_{i,n}$ is the distance between the two galaxies, and $C$ is a constant set to $-10.96$ so that galaxies with $\Theta_5 < 0$ represent isolated systems. The calculation is limited to galaxies within a distance range of 1 Mpc. For three galaxies (UGC00685, NGC6789, IC4662) no galaxies were identified within 1 Mpc; in these cases we assign a $\Theta_5$ value to be $-3.5$ for plotting purposes, which is lower than the values for any other galaxies in the sample. The second metric is the galaxy's distance to the nearest large (\mstar $>10^9$ \msun) galaxy (NLG). The NLG metric does not account for potential differences in virial radii of galaxies of similar stellar mass, but given the uncertainties in virial mass and radii of galaxies, the NLG metric is useful as it provides a straight-forward and uniform comparison across the sample.

\begin{deluxetable*}{l cccr | rrcc}
\tabletypesize{\footnotesize}
\label{tab:Obudget}
\tablecaption{Oxygen Budget}
\tablehead{
\colhead{Galaxy} & \multicolumn{4}{c}{log(M$_{\rm O}$/\mstar)} & \multicolumn{4}{c}{\% Oxygen retained} \\
\colhead{}	& \colhead{total} & \colhead{gas} & \colhead{stars}  & \colhead{dust} & \colhead{total} & \colhead{gas} & \colhead{stars}  & \colhead{dust} 
} 
\colnumbers
\startdata
WLM      & 5.93 $\pm$ 0.12 & 4.81 $\pm$ 0.10 & 3.78$_{-0.09}^{+0.17}$ & 3.67 & 8.90  $_{-3.00}^{+3.00}$ & 7.70  $\pm$ 2.80 & 0.70 $_{-0.20}^{+3.00}$ & 0.60 \\
UGC00685 & 6.14 $\pm$ 0.12 & 5.04 $\pm$ 0.03 & 4.74$_{-0.11}^{+0.10}$ & 4.21 & 13.20 $_{-3.80}^{+3.80}$ & 8.00  $\pm$ 2.30 & 4.00 $_{-1.50}^{+3.80}$ & 1.20 \\
NGC0625  & 7.02 $\pm$ 0.12 & 5.35 $\pm$ 0.12 & 5.59$_{-0.07}^{+0.15}$ & 3.81 & 6.00  $_{-1.90}^{+2.20}$ & 2.20  $\pm$ 0.80 & 3.80 $_{-1.20}^{+2.20}$ & 0.10 \\
NGC0784  & 6.90 $\pm$ 0.12 & 5.76 $\pm$ 0.05 & 5.48$_{-0.06}^{+0.09}$ & 5.24 & 13.40 $_{-3.80}^{+3.90}$ & 7.40  $\pm$ 2.20 & 3.80 $_{-1.20}^{+3.90}$ & 2.20 \\
NGC1569  & 7.23 $\pm$ 0.12 & 5.67 $\pm$ 0.05 & 5.46$_{-0.05}^{+0.13}$ & 4.88 & 4.90  $_{-1.40}^{+1.50}$ & 2.80  $\pm$ 0.80 & 1.70 $_{-0.50}^{+1.50}$ & 0.50 \\
NGC2366  & 6.65 $\pm$ 0.12 & 5.76 $\pm$ 0.05 & 4.64$_{-0.04}^{+0.06}$ & 5.00 & 16.30 $_{-4.80}^{+4.80}$ & 13.10 $\pm$ 3.90 & 1.00 $_{-0.30}^{+4.80}$ & 2.30 \\
UGC04305 & 6.71 $\pm$ 0.12 & 5.75 $\pm$ 0.10 & 4.75$_{-0.06}^{+0.09}$ & 5.34 & 16.50 $_{-5.20}^{+5.20}$ & 11.10 $\pm$ 4.00 & 1.10 $_{-0.30}^{+5.20}$ & 4.30 \\
UGC04459 & 5.56 $\pm$ 0.12 & 4.84 $\pm$ 0.09 & 3.59$_{-0.07}^{+0.08}$ & 3.97 & 22.80 $_{-7.40}^{+7.40}$ & 19.10 $\pm$ 6.60 & 1.10 $_{-0.30}^{+7.40}$ & 2.60 \\
UGC04483 & 5.22 $\pm$ 0.12 & 4.16 $\pm$ 0.03 & 3.08$_{-0.07}^{+0.11}$ & 2.65 & 9.80  $_{-2.80}^{+2.80}$ & 8.80  $\pm$ 2.50 & 0.70 $_{-0.20}^{+2.80}$ & 0.30 \\
UGC05139 & 6.09 $\pm$ 0.12 & 5.43 $\pm$ 0.05 & 4.07$_{-0.06}^{+0.08}$ & 4.39 & 24.80 $_{-7.30}^{+7.30}$ & 21.80 $\pm$ 6.50 & 1.00 $_{-0.30}^{+7.30}$ & 2.00 \\
UGC05364 & 4.67 $\pm$ 0.12 & 3.23 $\pm$ 0.05 & 2.59$_{-0.03}^{+0.07}$ & 1.15 & 4.60  $_{-1.30}^{+1.30}$ & 3.70  $\pm$ 1.10 & 0.80 $_{-0.20}^{+1.30}$ & 0.00 \\
SextansB & 5.82 $\pm$ 0.12 & 4.64 $\pm$ 0.05 & 3.76$_{-0.09}^{+0.14}$ & 3.31 & 7.80  $_{-2.30}^{+2.30}$ & 6.60  $\pm$ 2.00 & 0.90 $_{-0.30}^{+2.30}$ & 0.30 \\
NGC3109  & 6.39 $\pm$ 0.12 & 5.36 $\pm$ 0.07 & 4.58$_{-0.08}^{+0.14}$ & 4.26 & 11.60 $_{-3.60}^{+3.60}$ & 9.30  $\pm$ 3.00 & 1.60 $_{-0.50}^{+3.60}$ & 0.80 \\
SextansA & 5.77 $\pm$ 0.12 & 4.55 $\pm$ 0.09 & 3.55$_{-0.06}^{+0.11}$ & 3.16 & 6.90  $_{-2.30}^{+2.30}$ & 6.00  $\pm$ 2.10 & 0.60 $_{-0.20}^{+2.30}$ & 0.20 \\
UGC05666 & 7.09 $\pm$ 0.12 & 6.23 $\pm$ 0.05 & 5.24$_{-0.04}^{+0.09}$ & 5.46 & 17.60 $_{-5.10}^{+5.10}$ & 13.90 $\pm$ 4.10 & 1.40 $_{-0.40}^{+5.10}$ & 2.30 \\
NGC3738  & 6.99 $\pm$ 0.12 & 5.37 $\pm$ 0.06 & 5.79$_{-0.05}^{+0.12}$ & 4.27 & 9.00  $_{-2.60}^{+3.10}$ & 2.40  $\pm$ 0.70 & 6.40 $_{-1.90}^{+3.10}$ & 0.20 \\
NGC3741  & 5.40 $\pm$ 0.12 & 4.23 $\pm$ 0.03 & 3.39$_{-0.13}^{+0.09}$ & 3.53 & 9.10  $_{-2.60}^{+2.60}$ & 6.80  $\pm$ 1.90 & 1.00 $_{-0.40}^{+2.60}$ & 1.40 \\
UGC06817 & 5.57 $\pm$ 0.12 & 4.50 $\pm$ 0.02 & 3.61$_{-0.09}^{+0.11}$ & 1.66 & 9.70  $_{-2.70}^{+2.70}$ & 8.60  $\pm$ 2.40 & 1.10 $_{-0.40}^{+2.70}$ & 0.00 \\
NGC4068  & 6.30 $\pm$ 0.12 & 5.44 $\pm$ 0.10 & 4.65$_{-0.06}^{+0.11}$ & 4.61 & 18.10 $_{-5.90}^{+6.00}$ & 13.80 $\pm$ 5.00 & 2.30 $_{-0.70}^{+6.00}$ & 2.00 \\
NGC4163  & 5.89 $\pm$ 0.12 & 4.01 $\pm$ 0.14 & 4.28$_{-0.08}^{+0.11}$ & 0.97 & 3.80  $_{-1.20}^{+1.30}$ & 1.30  $\pm$ 0.60 & 2.40 $_{-0.80}^{+1.30}$ & 0.00 \\
CGCG269-049&5.25$\pm$ 0.12 & 4.12 $\pm$ 0.02 & 3.51$_{-0.17}^{+0.10}$ & 2.38 & 9.40  $_{-2.70}^{+2.70}$ & 7.40  $\pm$ 2.10 & 1.80 $_{-0.90}^{+2.70}$ & 0.10 \\
UGCA281  & 5.82 $\pm$ 0.12 & 4.72 $\pm$ 0.03 & 4.00$_{-0.16}^{+0.10}$ & 0.0 & 9.50  $_{-2.70}^{+2.70}$ & 8.00  $\pm$ 2.30 & 1.50 $_{-0.70}^{+2.70}$ & 0.00 \\
UGC07577 & 6.07 $\pm$ 0.12 & 4.71 $\pm$ 0.06 & 4.48$_{-0.06}^{+0.10}$ & 3.16 & 7.10  $_{-2.10}^{+2.10}$ & 4.40  $\pm$ 1.40 & 2.60 $_{-0.80}^{+2.10}$ & 0.10 \\
NGC4449  & 7.65 $\pm$ 0.12 & 6.59 $\pm$ 0.03 & 6.43$_{-0.05}^{+0.10}$ & 5.91 & 16.70 $_{-4.70}^{+4.90}$ & 8.90  $\pm$ 2.50 & 6.00 $_{-1.80}^{+4.90}$ & 1.80 \\
UGCA292  & 5.41 $\pm$ 0.12 & 4.18 $\pm$ 0.03 & 3.03$_{-0.10}^{+0.20}$ & 2.86 & 6.60  $_{-1.90}^{+1.90}$ & 5.90  $\pm$ 1.70 & 0.40 $_{-0.20}^{+1.90}$ & 0.30 \\
UGC08024 & 5.72 $\pm$ 0.12 & 4.99 $\pm$ 0.06 & 3.71$_{-0.08}^{+0.07}$ & 3.05 & 20.10 $_{-6.10}^{+6.10}$ & 18.90 $\pm$ 5.80 & 1.00 $_{-0.30}^{+6.10}$ & 0.20 \\
UGC08091 & 4.98 $\pm$ 0.12 & 3.80 $\pm$ 0.06 & 3.09$_{-0.07}^{+0.08}$ & 0.0 & 8.00  $_{-2.40}^{+2.40}$ & 6.70  $\pm$ 2.10 & 1.30 $_{-0.40}^{+2.40}$ & 0.00 \\
UGC08201 & 6.26 $\pm$ 0.12 & 5.19 $\pm$ 0.06 & 4.12$_{-0.07}^{+0.10}$ & 0.0 & 9.40  $_{-2.90}^{+2.90}$ & 8.60  $\pm$ 2.70 & 0.70 $_{-0.20}^{+2.90}$ & 0.00 \\
UGC08508 & 5.53 $\pm$ 0.12 & 4.26 $\pm$ 0.07 & 3.43$_{-0.07}^{+0.08}$ & 2.69 & 6.30  $_{-2.00}^{+2.00}$ & 5.40  $\pm$ 1.70 & 0.80 $_{-0.30}^{+2.00}$ & 0.10 \\
UGC08638 & 5.84 $\pm$ 0.12 & 4.44 $\pm$ 0.05 & 4.25$_{-0.09}^{+0.11}$ & 3.43 & 7.00  $_{-2.10}^{+2.10}$ & 4.00  $\pm$ 1.20 & 2.60 $_{-0.90}^{+2.10}$ & 0.40 \\
UGC08651 & 5.50 $\pm$ 0.12 & 4.53 $\pm$ 0.04 & 3.46$_{-0.08}^{+0.08}$ & 3.09 & 12.00 $_{-3.50}^{+3.50}$ & 10.70 $\pm$ 3.10 & 0.90 $_{-0.30}^{+3.50}$ & 0.40 \\
NGC5253  & 7.19 $\pm$ 0.12 & 5.40 $\pm$ 0.10 & 5.66$_{-0.03}^{+0.08}$ & 4.12 & 4.70  $_{-1.40}^{+1.50}$ & 1.60  $\pm$ 0.60 & 3.00 $_{-0.80}^{+1.50}$ & 0.10 \\
UGC09128 & 5.10 $\pm$ 0.12 & 4.14 $\pm$ 0.05 & 2.96$_{-0.05}^{+0.07}$ & 2.67 & 12.20 $_{-3.60}^{+3.60}$ & 11.10 $\pm$ 3.30 & 0.70 $_{-0.20}^{+3.60}$ & 0.40 \\
UGC09240 & 5.84 $\pm$ 0.12 & 4.73 $\pm$ 0.03 & 3.87$_{-0.07}^{+0.09}$ & 4.08 & 10.70 $_{-3.00}^{+3.00}$ & 7.90  $\pm$ 2.20 & 1.10 $_{-0.30}^{+3.00}$ & 1.80 \\
IC4662   & 6.43 $\pm$ 0.12 & 4.88 $\pm$ 0.04 & 5.13$_{-0.03}^{+0.11}$ & 3.79 & 8.10  $_{-2.30}^{+2.60}$ & 2.80  $\pm$ 0.80 & 5.10 $_{-1.50}^{+2.60}$ & 0.20 \\
NGC6789  & 5.90 $\pm$ 0.12 & 4.20 $\pm$ 0.05 & 4.52$_{-0.12}^{+0.10}$ & 3.38 & 6.50  $_{-2.20}^{+2.10}$ & 2.00  $\pm$ 0.60 & 4.20 $_{-1.70}^{+2.10}$ & 0.30 \\
IC5152   & 6.50 $\pm$ 0.12 & 4.82 $\pm$ 0.07 & 5.23$_{-0.07}^{+0.12}$ & 4.01 & 7.90  $_{-2.40}^{+2.60}$ & 2.10  $\pm$ 0.70 & 5.40 $_{-1.70}^{+2.60}$ & 0.30 \\\enddata
\tablecomments{Cols 2$-$5 list the mass of oxygen produced by stellar nucleosynthesis and the mass estimated to be in the gas, stars, and dust components of the galaxies. Cols 6$-$9 provide the percent of oxygen retained overall in the galaxies as well as the breakdown of retention in the gas, star, and dust. 
}
\end{deluxetable*}

The top panel of Figure~\ref{fig:orf_env} shows that the galaxies residing in denser environments as measured by the tidal index, $\Theta_5$, retain a greater fraction of oxygen. The exceptions are galaxies with low gas fractions, as shown by the color-coding of the points. Since the majority of oxygen is found in the ISM in low-mass galaxies, this just reflects that the galaxies with less \hi\ also have lower oxygen retention. 

The bottom panel of Figure~\ref{fig:orf_env} shows that galaxies at larger distances ($>1$ Mpc) from a more massive (\mstar $>10^9$ \msun) neighbor have consistently low oxygen retention fractions ($<15$\%). All of these galaxies also have low tidal indices ($\Theta_5 < 0$) indicating they reside in less dense environments, with the exception of one galaxy, NGC~4163 ($\Theta_5 = 1.6$). 

However, when residing within 1 Mpc of a more massive system, the picture becomes more complex and there is significant scatter in the retention fractions, indicating environment impacts the amount of oxygen retained, recycled, or even accreted to galaxies. On closer inspection, we note that the density of the environment may be driving the scatter: of the 9 systems with retention fractions consistent with more isolated galaxies (i.e., $<$15\%), 6 galaxies (67\%) have low tidal indices ($\Theta_5 < 0$). In contrast, nearly all of the galaxies (7/8 or 88\%) in similar proximity ($<$1 Mpc) to a more massive neighbor that are also experiencing greater tidal field strengths ($\Theta_5 > 0$; highlighted by the dashed box) are the only galaxies with higher retention fractions ($>15$\%). 

The trend that higher retention fractions may be associated with both closer proximity to a more massive neighbor and in higher density galactic environments is tentative and based on a small sample size. If this trend is confirmed with a larger sample, it could imply a few different scenarios: (i) gas inflows into galaxies residing in more complex environments may be pre-enriched by metals expelled by nearby galaxies \citep[which would indicate metal and gas transfer between galaxies;][]{Hafen:2020_fateCGM}; (ii) there is a greater degree of recycling of ejected metals back to the disks of galaxies in these environments \citep{Armillotta:2016uq}; or (iii) both. 

In the first scenario, the local IGM or, in cases where a galaxy resides in a galaxy group, the inter-group medium, may be more highly `polluted' by metals being ejected or stripped via hydrodynamical interactions with galaxies and/or gas in the surroundings. A comprehensive study of the \hi\ content in low-mass galaxies within the Local Group suggests that the Local Group medium likely played a role in removing the gas from these systems \citep{Putman2021}. There is evidence that the metal content in the CGM of galaxies in group environments, traced by Mg~\textsc{ii}, is higher and more extended compared to isolated galaxies at z$\sim$1, suggesting ram pressure stripping in denser environments may help to `lift' metals away from galaxy disks \citep[e.g.,][and references therein]{Dutta2023, Fumagalli2024}, which are then available to be accreted by other nearby galaxies in the group.  Other studies find suppressed CGM contents of \hi\ and \CIV\ in dense environments at low redshift \citep{Burchett:2016aa, Burchett:2018aa}, suggesting that the material, if present, may exist in a highly ionized phase. Note that, in this scenario, greater gas accretion could also be possible, which would increase the gas fractions and effective yields. If this is occurring, it is likely a stochastic process impacting only a subset of galaxies based on $f_{gas}$ and the scatter in the amount of \hi\ within 4.4$\alpha$ of the galaxies that are within 1 Mpc of a NLG.

In the second scenario, the surrounding medium in denser environments may have higher temperatures and/or densities resulting in a higher ambient pressure. The higher pressure could reduce the efficiency of metal expulsion from galaxy halos and increases the probability of metals being recycled onto the galactic disks. Observational evidence supports this idea: galaxies with at least one \mstar $=10^{9.5}$ \msun\ neighbor were reported to have \ion{O}{6} covering fractions in their CGM of $1.5-5$ times greater than that found in galaxies without such neighbors \citep{Tchernyshyov2022}, suggesting a greater degree of metal recycling is indeed possible in denser environments. 

It is also possible that the outer regions of the dark matter halos in denser environments might be stripped or truncated, creating a more concentrated and deeper gravitational potential which is more effective at retaining or recycling metals into a galaxy's disk. However, as shown in the bottom panel of Figure~\ref{fig:orf_env}, the \hi\ disks in galaxies within 1 Mpc of their nearest largest neighbor do not appear to be truncated, regardless of tidal index $\Theta_5$, with the majority having $>$25\% of their \hi\ flux outside 4.4 scale lengths. In fact, a number of the galaxies in regions of stronger tidal field strength have some of the most extended \hi\ disks (see Paper I). This would suggest that any stripping is modest, emphasizing that while there is some diversity in the environments probed, overall we are sampling only a narrow range of parameters and the GLOW sample are overall considered field galaxies.

\section{Discussion}\label{sec:discuss}
In this section, we begin by exploring whether the missing oxygen atoms reside in the CGM or have been lost from the galaxy halos (\S~\ref{sec:cgm}). We then put our results into greater context by expanding the analysis to a larger mass range of galaxies ($10^5 <$ \mstar\ $<10^{11.5}$) using results from the literature (\S~\ref{sec:compare_empirical}), and comparing our results with predictions from hydrodynamical simulations (\S~\ref{sec:compare_models}). Finally, we use a regulator-type model of galaxy evolution to explore different wind models and pinpoint the key parameters that may explain the mismatch between our empirical results and predictions from simulations (\S~\ref{sec:regulatory_model}).

\subsection{Where are the Missing Oxygen Atoms? Constraints on the CGM from the Literature}\label{sec:cgm}
The GLOW galaxies have retained only a small fraction of the oxygen produced by stellar nucleosynthesis within the stellar bodies and main gaseous disks. It is an open question what fraction of the expelled atoms reside in the CGM vs.\ what fraction has been removed entirely from the galactic gravitational potential wells and now reside in the larger intergalactic medium (IGM). 

There have been several attempts to constrain the metal content in the CGM surrounding low-mass galaxies over a large range in stellar mass ($10^{6.5} - 10^{9}$ \msun). These studies use UV absorption spectroscopy with the HST Cosmic Origins Spectrograph (COS) and line-of-sight background quasars both in local galaxies \citep[i.e., within several Mpc;][]{Zheng2019, Zheng2020, Zheng2024, Fox2026} and at low redshift \citep[e.g., $z\ltsimeq0.6$;][]{Bordoloi2014, Liang2017, Johnson2017, Tchernyshyov2022, Mishra2024} to search for metals. The analyses, however, are hampered by the limited number of UV sightlines, the random distributions of impact parameters per system, and, critically, the low density of the CGM gas in dwarfs, which is difficult to detect given the sensitivity limits of the COS instrument; these factors can conspire to return low detection rates of metals. Significant effort on modelling the (mostly) non-detections aid interpretations \citep[e.g.,][]{Zheng2024}, but ambiguity remains as to whether the low detection rates are due to limitations in the data or the actual lack of metals in the CGM of low-mass galaxies. 

In the context of this study focusing on oxygen, there is an additional challenge: the species probed in the CGM of the nearest of galaxies are typically the low-ionization states of carbon (\ion{C}{2}, \ion{C}{4}) and silicon (\ion{Si}{2}, \ion{Si}{3}, \ion{Si}{4}) dictated by the wavelength range of COS\footnote{\ion{O}{6} falls outside the COS high-resolution grating wavelength range for $z\ltsimeq0.1$.}. Using these species as a tracer of the oxygen content in the CGM is not straightforward. First, the elements are forged via different processes: while oxygen is produced nearly entirely by CCSNe, carbon is also produced by asymptotic giant branch (AGB) stars and silicon is produced by Type I supernovae. The different origins of carbon and silicon change the yields of these species relative to oxygen that depends on the detailed SFHs of galaxies and, in the case of carbon, the still uncertain modelling of AGB stars \citep[see, e.g.,][]{Kobayashi2006, Kobayashi2020}. 

Second, the different production mechanisms of the elements also imply different distribution channels and expulsion rates. Oxygen transport is powered by supernovae physics which can accelerate oxygen to high-velocities with the possibility of escape from the main stellar body whereas much of the carbon transport is driven by AGB stellar pulsations and stellar winds which results in lower velocities and a lower probability of expulsion. Indeed, measured carbon to oxygen ratios have been shown to vary by nearly 0.5 dex for a given 12$+$log(O/H) value \citep{Berg2019}. These authors find that the C/O ratio is very sensitive to both the detailed SFH and to supernova feedback, and that a declining C/O relationship is seen with increasing baryonic mass due to increasing effective oxygen yields. Silicon transport driven by Type I supernovae occurs at different rates than the Type IIs which also needs to be taken into account. 

The above uncertainties mean that converting measurements (or, in most cases, the observed upper limits) of C and Si retained in the CGM to robust constraints on the O mass in the CGM is difficult. When moving from $z\sim0$ to low-redshift ($z\ltsimeq0.6$), a higher ionization state of oxygen (\ion{O}{6}) falls within the COS wavelength range, overcoming the inherent challenges in comparing different elements, but now introduces the inconsistency that the content and structure of the CGM is expected to evolve over the more than $\sim5$ Gyr in time from $z\sim0.6$ to the present-day, which is also not well-constrained.

With these limitations in mind, we proceed cautiously with a comparison of the oxygen mass estimated to reside in the CGM from low-mass galaxies from previous work and the \ORF\ values we calculate, and use constraints on other species to inform our conclusions. 

The most comprehensive results specifically on oxygen are from \citet{Tchernyshyov2022} who reported the median mass of the \ion{O}{6} ion in star-forming galaxies in stellar mass bins relevant to GLOW. Specifically, they estimate log(M$_{\rm O{\sc VI}}$/\msun) $=5.2\pm0.2$ for galaxies with \mstar $=10^{7.8-8.5}$ \msun, $5.6^{+0.1}_{-0.4}$ for \mstar $=10^{8.5-9.2}$ \msun, and $6.1\pm0.1$ for \mstar $=10^{9.2-9.6}$ \msun; their sample spans a redshift range of $z\sim0.1-0.6$. \citet{Johnson2017} estimate a similar value of log(M$_{\rm O{\sc VI}}$/\msun) $=5.5$ \msun\ for galaxies with \mstar\ $=10^{8.4}$ \msun\ in the redshift range of $z=0.1-0.3$. \citet{Mishra2024} used the sample from \citet{Johnson2017} and expanded the number of dwarfs with the requisite observations by pulling observations from a larger volume (i.e., a redshift range of $z=0.07-0.73$).  These authors calculate a comparable value of log(M$_{\rm O{\sc VI}}$/\msun) $=5.6\pm0.2$ \msun\ within a virial radius for galaxies with a mean \mstar\ $=10^{8.3}$ \msun, and a similar value of log(M$_{\rm O{\sc VI}}$/\msun) $=5.5\pm0.2$ \msun\ residing between $1-2$ virial radii of the systems. Values from these fours studies are overplotted in Figure~\ref{fig:orf_components}. With the significant caveat that the \ion{O}{6} measurements from these studies are at redshifts of $z\sim0.1-0.7$, the estimated masses of these highly ionized atoms represent $\sim$3-10\% of the oxygen production in GLOW galaxies at $z=0$.

The mass of \ion{O}{6} in the CGM in very local galaxies, including galaxies that are part of the GLOW sample, has been estimated using empirical constraints (mostly non-detections) of C and Si and then modelling the expected oxygen mass \citep{Zheng2024}. These authors predict the mass in the \ion{O}{6} in the CGM to be $10^{5.1}$ \msun\ for a galaxy with \mstar\ $=10^{8.3}$; the modelling assumed that a composite of the observations from galaxies with \mstar\ ranging from $=10^{6.5} - 10^{9.5}$ \msun\ is representative of the median mass of the sample (\mstar\ $=10^{8.3}$ \msun; M$_{200} = 10^{10.9}$ \msun). We overplot this value in Figure~\ref{fig:orf_components} for comparison to the oxygen mass calculations for the GLOW sample and note that the modeled mass of \ion{O}{6} corresponds to $\sim$3\% of the oxygen produced by a galaxy with \mstar\ $=10^{8.3}$ \msun. 

Translating the mass of a single oxygen ionic species, namely \ion{O}{6}, to a total oxygen mass in the CGM is highly uncertain.  Even with caveats such as (likely) non-uniform metallicity, collisional/photoionization equilibrium conditions, and volume filling factors throughout the CGM, the necessary ionization corrections vary widely depending on the local physical conditions. This is especially true for \ion{O}{6}, which traces at most $\sim20\%$ of the total oxygen within a narrow temperature range around $10^{5.5}$ K under collisional ionization equilibrium and $>20\%$ at very low densities $n_H<10^{-4}$ cm$^{-3}$ under photoionization equilibrium \citep[e.g.,][]{Oppenheimer2012}. Given that the virial temperature of galaxies in the GLOW mass range is expected to be $\lesssim10^{5.2}$ K and the baryonic density distribution in the CGM is uncertain, photoionization is the more likely mechanism.  If, on average, 20\% of the oxygen is in the \ion{O}{6} state, based on the results discussed above, this would imply the CGM reservoirs of dwarfs contain only $5-35$\% of the oxygen produced by nucleosynthesis. The results from \citet{Mishra2024} suggest that some of the missing oxygen may reside in the IGM in close proximity to the galaxies (i.e., within $1-2$ virial radii). 

While the lower ionization states of oxygen are not observable in the CGM, studies of the low and intermediate ions of carbon and silicon detectable in the UV estimate have been done, with varying results. From two of the previous studies mentioned, roughly 10\% of carbon and silicon reside in the cool and warm phases \citep[e.g.,][]{Johnson2017, Zheng2020}. Hydrodynamical simulations predict similarly lower percentages in the CGM of low-mass galaxies in the cool and warm phases ($T\lesssim10^{5.5}$), with each phase containing roughly 10\% of the total metals produced by star formation \citep[e.g.,][]{Muratov2017, Christensen2018, Piacitelli2025}. Note, however, that the simulations predict that low-mass galaxies harbor a much larger reservoir of oxygen in the ISM than observed (see Section~\ref{sec:compare_models} and Figure~\ref{fig:orf_models}). If the amount retained in the disks were to be reduced, this could result in a larger amount of oxygen residing in the CGM.

A lower fraction of $\sim0.5$\% of metals residing in the cool CGM was reported in a study that measured the silicon content in the CGM of the nearby galaxy Sextans~B \citep{Fox2026}, a galaxy also included in GLOW. \citet{Fox2026} reports that the CGM of Sextans~B contains $5.5\times10^2$ \msun\ of silicon within an 8 kpc radius of the galaxy disk \citep{Fox2026}. The amount  of total silicon detected in the CGM around Sextans~B is within a factor of two of the amount of oxygen we report to be retained in the stars of this same galaxy. Assuming the density profile declines with radius in the CGM, this result suggests that there is only a small fraction of metals produced by star formation that reside in the cool phase of the inner CGM of this galaxy.

In an accompanying work to the present study (J. N. Burchett et al., in preparation), we analyze three sightlines probing the CGM of Sextans A and detect lines from several ions, notably \ion{C}{2} and \ion{C}{4}.  We find the metallicity of the extended diffuse gas to be $Z/Z_\odot \sim 0.02$, similarly low to that derived by \citet{Fox2026} for Sextans B. Using the density profile reported by \citet{McQuinn2022} and this metallicity, we find that 80-100\% of the missing carbon (distributed among its various ionization states) from Sextans A may be retained in the CGM.  This modeling assumes on a large volume filling factor but does suggest Sextans A retains a significant fraction of its carbon produced and expelled from the ISM. 

In summary, the mass of oxygen in \ion{O}{6} detected in the CGM is less than the mass of oxygen we measure in the disks of the GLOW galaxies, as shown in Figure~\ref{fig:orf_components}, and accounts for $1-7$\% of the oxygen produced via stellar nucleosynthesis. Assuming a ionization correction factor of 0.2, the total oxygen budget in the CGM may be between $5-35$\%. These empirical values could be quite different if the ionization correction for oxygen used on the observations does not apply in the CGM of dwarf galaxies due to nonequilibrium processes. As the lower ionization states of oxygen are not detectable, we also examined results from the literature on the low and intermediate ions of carbon and silicon. The results do vary, but overall suggest the cool and warm phases in the CGM harbor only a fraction of the metals lost from the galaxies. We conservatively conclude that the majority of oxygen in the low-mass galaxies remains unaccounted for; it is still an open question as to whether oxygen missing from the disks of low-mass galaxies has been expelled from the gravitational wells and lost to the IGM.

\subsection{Retention Fractions from \mstar\ $\sim10^5$ to $10^{11.5}$ \msun}\label{sec:compare_empirical}
\begin{figure*}
\includegraphics[width=0.9\textwidth]{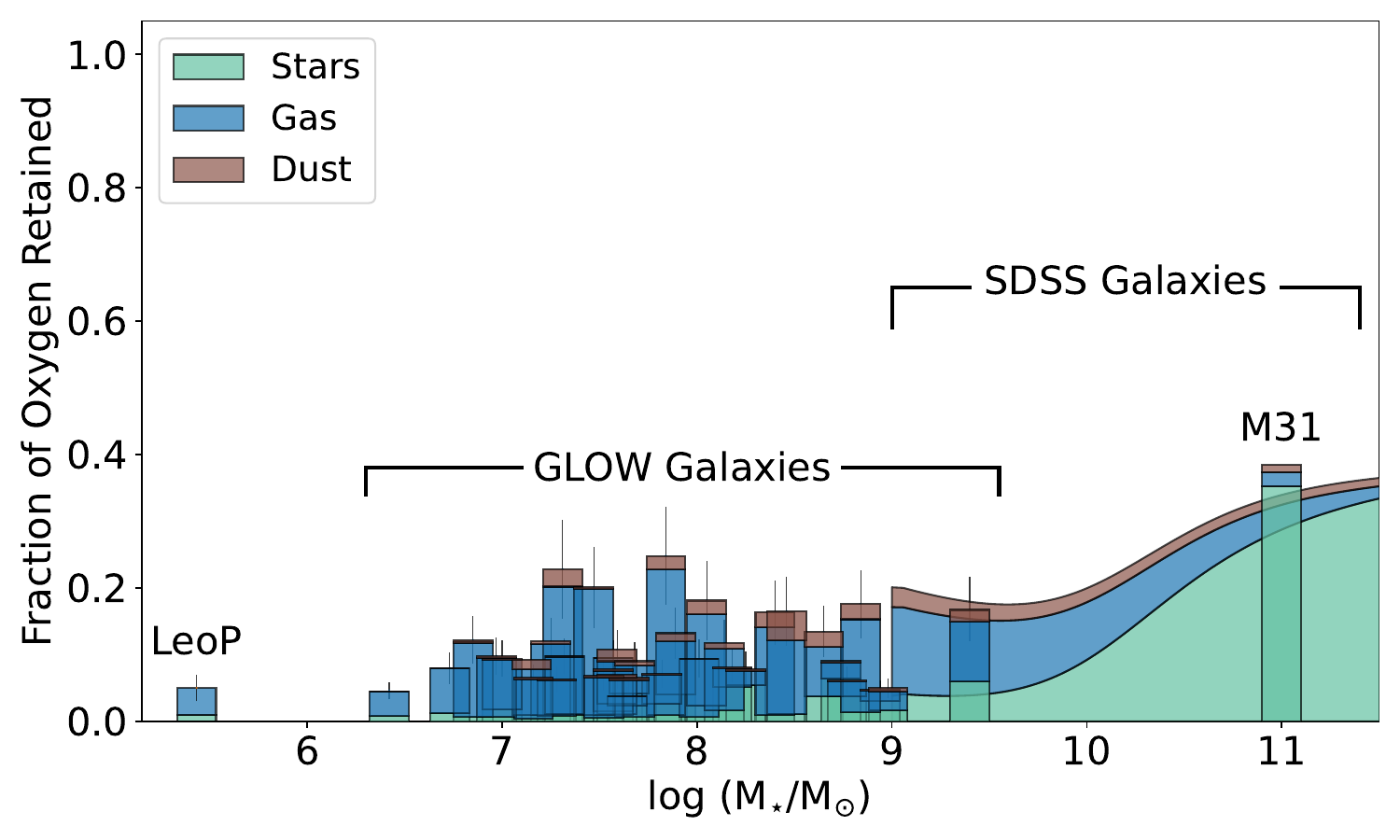}
\caption{Fraction of oxygen retained in the stars (green), gas (blue), and dust (brown) for each of the individual GLOW galaxies as a function of stellar mass. Also shown are similar measurements from the very low-mass galaxy Leo~P \citep{McQuinn2015b, McQuinn2024}, the metal retention fraction in the stars, gas, dust in disk of the massive spiral M31 \citep[][]{Telford2019}, and estimates of oxygen retained in the stars, gas, dust of SDSS galaxies based on scaling relations \citep{Peeples2014}.}
\label{fig:orf_parsec}
\end{figure*}

Figure~\ref{fig:orf_parsec} presents the \ORF\ results for the GLOW sample as function of stellar mass, now color-coding the fraction of oxygen in each component (i.e., stars, gas, dust) on a galaxy-by-galaxy basis. In addition, we expand the range in stellar mass probed by adding results from the literature. Specifically, we show the retention fraction from the very low-mass galaxy, Leo~P \citep{McQuinn2015b, McQuinn2024}, which is a prototype for the analysis presented here, and the metal retention fraction derived from spatially resolved analysis of our neighboring spiral galaxy M31 using a similar approach \citep{Telford2019}. Leo~P provides an anchor at the low-mass end of our sample, and has retained 5\% of its oxygen. Although smaller retention fractions of $1-3$\% have been reported for low-mass (\mstar\ $\sim10^{5.5}-10^{7.5}$ \msun) satellites of the Milky Way, these satellite galaxies have lost their gas, presumably removed via significant environmental processing (e.g., ram pressure and tidal stripping), and the metals retained reside solely in the stars \citep{Kirby2011}. M31 is a spiral galaxy orders of magnitude more massive than the GLOW sample \citep[\mstar\ $=1.03^{+0.23}_{-0.17} \times 10^{11}$ \msun;][]{Sick2015}. \citet{Telford2019}  tallies a sum total of 38\% of the metals have been retained across all disk components (stars:\,35.3\%, gas:\,2\%, dust:\,1.2\%). The considerably higher retention of metals in M31 is expected over low-mass galaxies given the deeper potential. Note that although the M31 analysis was based on data from the PHAT survey footprint \citep{Dalcanton2012} that covered approximately half of the main stellar disk, results were extrapolated to be representative of M31's full stellar disk by assuming axial symmetry. 

We also reproduce constraints for SDSS for galaxies with \mstar\ of $10^{9}-10^{11.5}$ \msun\ from \citet{Peeples2014} where, instead of analyzing individual galaxies, the trend in oxygen retention fractions was constrained using galaxy scaling relations and trends in galaxy metallicity measurements. While the results from this study were extrapolated down to \mstar\ $= 10^{8.5}$ \msun\ in \citet{Tumlinson2017}, we limit our comparison to the mass range of the data used in the scaling relations for the derivations in \citet{Peeples2014} to avoid additional uncertainties. We focus solely on the {\em oxygen} retention fractions reported in \citet{Peeples2014}, which decrease by a factor of two from the higher to lower mass galaxies. Note that this is in contrast with the {\em metal} retention fractions reported in \citet{Peeples2014} which showed a surprisingly flat retention fraction as a function of stellar mass over the same stellar mass range. 

From Figure~\ref{fig:orf_parsec}, at the low-mass end, our values are consistent with expectations anchored by Leo~P with \mstar\ $=2\times10^5$ \msun\ \citep{McQuinn2015b, McQuinn2024} which was derived using a similar approach. The higher mass end of our sample overlaps in galaxy mass with the lower mass regime studied by \citet{Peeples2014} (e.g., log(\mstar/\msun) $\sim9$). As the two approaches used disparate data sets, methods, and assumptions, these overlapping regions provide a useful comparison and check on our results. 

Indeed, remarkably, there is general agreement between the studies, although our retention fractions are lower than the values from \citet{Peeples2014}. Specifically, we find that the GLOW galaxies with \mstar\ $> 5\times10^8$ \msun\ have a mean \ORF\ of 10\%, compared with the mean value of 20\% reported for \mstar\ $= 10^9$ \msun\ by \citet{Peeples2014}. The possible cause of this factor of two difference is difficult to discern and we cannot rule out that different assumptions adopted between the studies may conspire to cancel or amplify the impacts. For example, the work from \citet{Peeples2014} is based on a statistical analysis and relies on scaling relations (e.g., the gas masses of galaxies are based on adjusting the stellar masses using an empirically fit function of stellar-to-gas ratios in galaxies from a variety of measurements from the literature), whereas our results are based on measurements of individual galaxies in all phases. If \citet{Peeples2014} had assumed lower gas fractions for the lower mass galaxies in the SDSS study, their retention fractions would decrease, bringing their results into rough agreement with our reported values. On the other hand, \citet{Peeples2014} adopted a slightly higher oxygen yield for the stars, namely 0.015 vs.\ our value of 0.01. If we adjust our calculations by using this higher yield value, the retention fractions we measure decrease by $1-10$\% with a mean value of 4\%, increasing the discrepancies with the SDSS study. 

Note also that, in contrast to the analysis of the GLOW galaxies, Leo~P, and M31, the study by \citet{Peeples2014} did not include the oxygen mass locked in stellar remnants in the stellar component, which can be appreciable in more massive galaxies. Thus, the fraction of oxygen in the stellar components should be considered lower limits compared to the values reported for the individual galaxies in GLOW, Leo~P, and M31. When comparing the upper end of \citet{Peeples2014} with the measurement from M31, a combination of these variables could also increase the retention fraction based on the scaling relations at the upper mass end and bring these two studies into even closer agreement. 

There are a few main takeaways from Figure~\ref{fig:orf_parsec}. First is the steady decline in the oxygen retention fraction as a function of galaxy mass: 38\% at \mstar\ $\sim10^{11}$ \msun\ for M31, $\sim$25\% at the high-mass end based on scaling relations, $<15$\% for low-mass ($10^7-10^8$ \msun) galaxies, and 5\% for \mstar\ $<10^6$ \msun. Second is the clear trend on where the oxygen resides within a galaxy: at higher masses, the oxygen in stars clearly dominate the budget \citep[even without the inclusion of stellar remnants in][]{Peeples2014}. As one to moves to lower masses (\mstar\ $<10^{10}$ \msun) the oxygen budget in the gas increases and overtakes that of the stars. While the study from \citet{Peeples2014} assumed a constant gas fraction for the galaxies, the GLOW results are based on measurements from individual galaxies; this more nuanced and detailed view into the gas vs.\ star components definitively confirms this trend and extends it to much lower masses. Finally, when also considering empirical constraints on the \ion{O}{6} mass in the CGM (Section~\ref{sec:cgm}), it remains an open question on whether the majority of the oxygen produced by stars in low-mass galaxies is retained by the galaxies in their CGM, or lost to the IGM.

\subsection{Comparison with Models}\label{sec:compare_models}
The production, distribution, and retention of metals in galaxies have been examined from a theoretical perspective via a number of high-resolution simulations across of a range of galaxy and halo masses \citep[e.g.,][]{Kobayashi2007, Taylor2020, Shen2014, Muratov2017, Christensen2018, Piacitelli2025}. Here, we focus on simulations that include quantitative predictions for the low-mass galaxy regime that break down the fraction of metals retained in individual galaxy components such that they can be readily compared with our calculations. 

Specifically, we consider the works of \citet{Muratov2017, Christensen2018, Piacitelli2025} and compare the predictions with our constraints from GLOW.  Despite being the most directly comparable to our results, caution is warranted in interpreting the comparisons as there are differing core assumptions employed between our empirical analysis and the theoretical frameworks including: the adopted nucleosynthetic yields, the upper mass limit for CCSNe, the return fraction of material per stellar generation (R), different definitions of the ISM extent, and, across the simulations, different implementations of the sub-grid physics responsible for driving stellar feedback, galactic outflows, and metal loss. While a strict, quantitative comparison of values in a given mass range can be difficult, overall trends between observations and simulations should be robust. Indeed, significant offsets between predictions and our empirical measurements can help discern between sub-grid baryon physics prescriptions and, in particular, feedback physics, used in the different simulations. 

\begin{figure}
\includegraphics[width=0.45\textwidth]{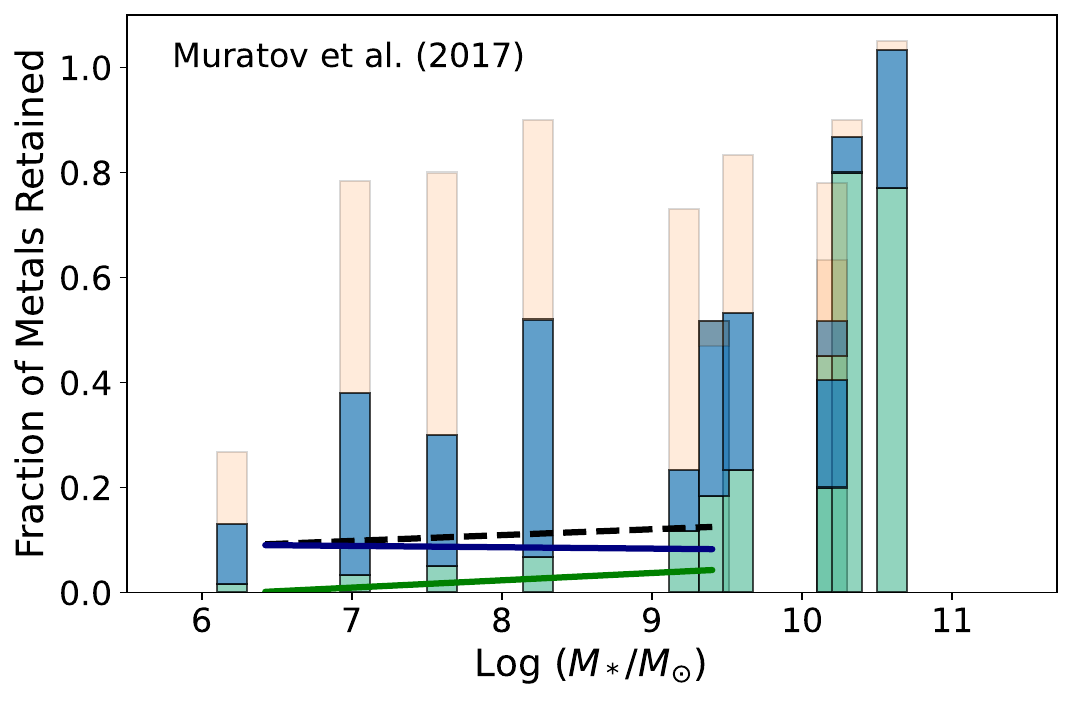}
\includegraphics[width=0.45\textwidth]{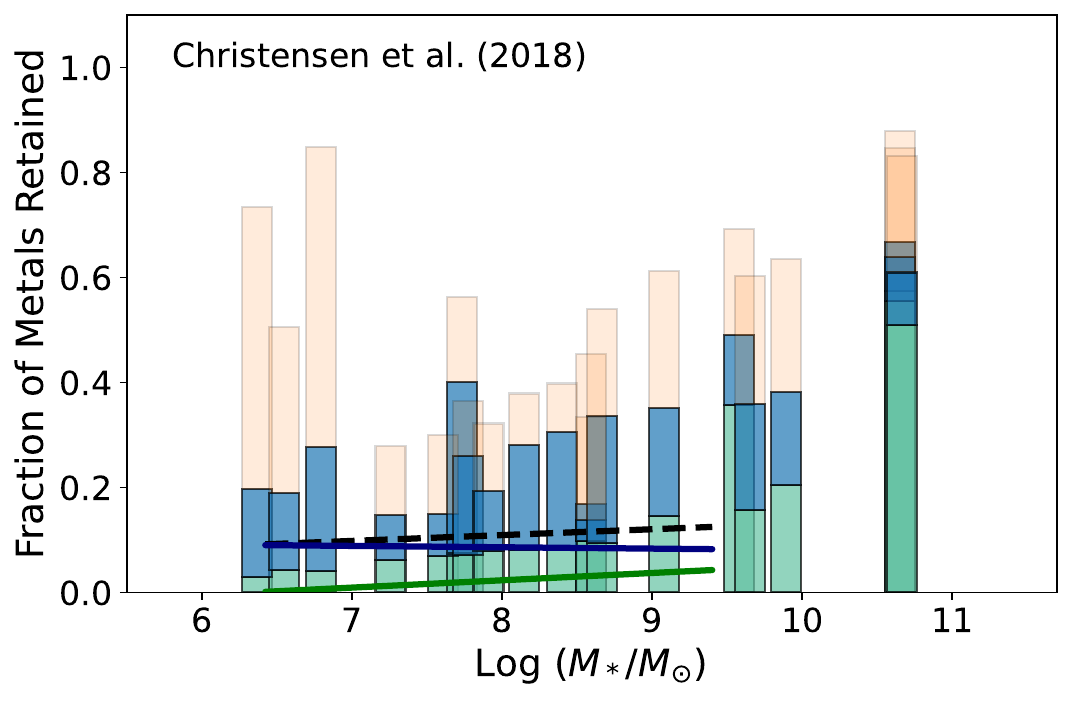}
\includegraphics[width=0.45\textwidth]{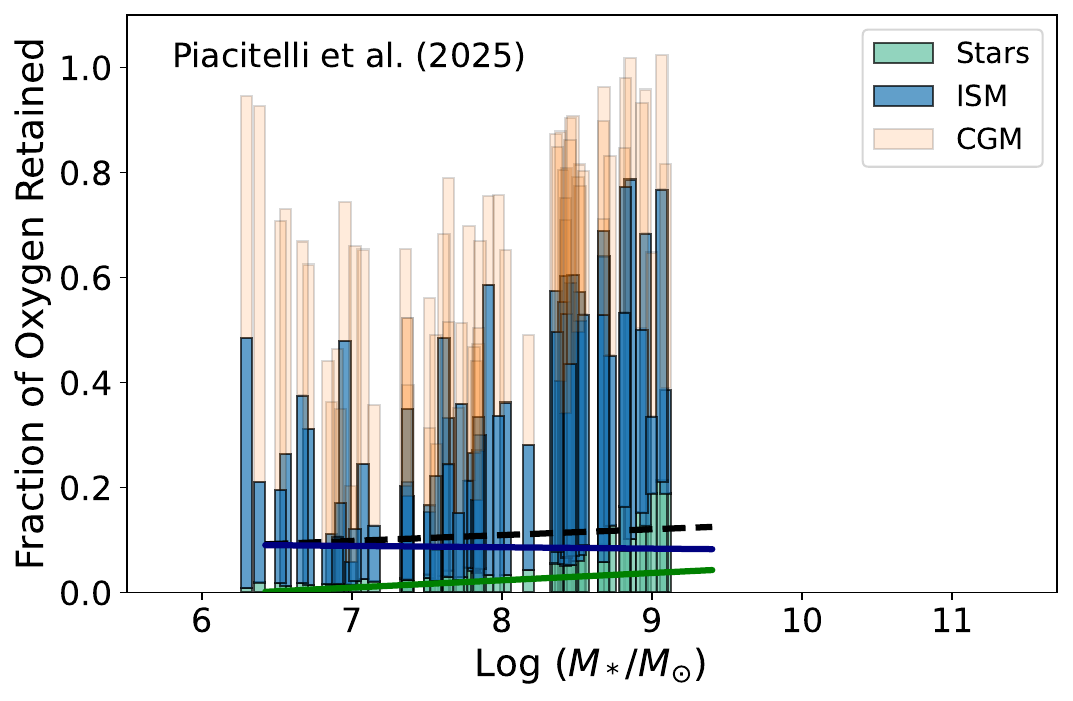}
\caption{Percent of metals (top 2 panels) or oxygen (bottom panel) retained in the stars, ISM, and CGM in three separate hydrodynamical simulations \citep{Muratov2017, Christensen2018, Piacitelli2025}. For comparison with GLOW, we also overplot on each panel solid green and blue lines representing the best-fit lines to fraction of oxygen retained in stars and the ISM of the GLOW galaxies, respectively, and a black dashed line as the best-fit line to the total oxygen retained (see Figure~\ref{fig:orf_properties}). Note that the results from \citet{Piacitelli2025} have been modified to show the percent of oxygen retained, rather the total metals originally reported in that work.}
\label{fig:orf_models}
\end{figure}

In Figure~\ref{fig:orf_models}, we reproduce results from smoothed particle hydrodynamic simulations, including \citet{Muratov2017} (top), \citet{Christensen2018} (middle), and from \citet{Piacitelli2025} (bottom); note that the results from \citet{Piacitelli2025} have been modified to reflect the oxygen retained rather than total metals. Similar to our measurements shown in Figure~\ref{fig:orf_parsec}, the predictions are for individual simulated galaxies and we adopt a similar color-scheme used for the observations, but note that blue includes both gas and dust and light orange is added to show the metal content in the CGM component. Overplotted on each is the mean retention fraction from stars (green solid line), gas and dust (blue solid line), and the total (black dashed line) of the GLOW galaxies (from Figure~\ref{fig:orf_properties}), which can be compared to the stars (green bars) and ISM (blue bars) in the simulation results. Note that the simulations from \citet{Muratov2017} and \citet{Christensen2018} report the overall metal retention fractions, while those shown from \citet{Piacitelli2025} are based solely on the oxygen fractions. There are subtle differences introduced by following metals vs.\ oxygen, although these differences should be small relative to the overall trends. Indeed, a comparison of the distribution of metals vs.\ oxygen fractions from \citet{Piacitelli2025} shows negligible difference in the stellar component, and a shift of only a few percent of the oxygen retained in the ISM into the CGM relative to the total metal framework (likely due to SN~II being the dominant production mechanism of oxygen). 

From Figure~\ref{fig:orf_models}, the simulations results from \citet{Muratov2017} predict higher retention fractions than observed by a factor of $\sim2-4$ across the mass regime of GLOW. At the higher mass end, the comparison becomes more difficult due to the significant jump in predictions (e.g., from $\sim$40\% to 100\% over the small mass range of \mstar\ $\sim10^{10.2}-10^{10.6}$), but all are higher than reported by \citet{Peeples2014} and higher than the 38\% for the even more massive galaxy M31 from \citet{Telford2019}. Looking at overall trends, these simulations do show an increase in retention fractions moving from lower to higher masses, although the sample size is small and there is significant variability. 

The results from \citet{Christensen2018} are a closer match to the empirically determined retention fractions, both in the low-mass regime of GLOW and for the higher mass galaxies studied by \citet{Peeples2014} and \citet{Telford2019}, but are still higher overall. While a detailed comparison for more massive galaxies is not the focus of our analysis, \citet{Christensen2018} noted these the differences in the retention fractions could be due to the different assumptions used in calculating the total metals produced by nucleosynthesis, which has a significant impact on the metal budget. On the other hand, our results from individual galaxies generally agree but are somewhat lower than the retention fractions from \citet{Peeples2014} in the overlapping mass regime, suggesting there may be other assumptions in the simulations to consider. Focusing on the lower-mass galaxy regime, the simulations from \citet{Christensen2018} are mostly within a factor of 2 of that measured from the GLOW galaxies, and predict similar scatter between galaxies. 

Finally, the predictions from \citet{Piacitelli2025}, which is the largest sample by far, include a few simulated galaxies that are a closer representation to GLOW galaxies, although there is significant scatter in the predictions and the majority of galaxies show retention fractions that are a factor of few higher than we observe. Note, also, the large overlap between the fraction of metals lost from the sample of \citet{Piacitelli2025} and the results from  \citet{Christensen2018}. 

In sum, all three simulations qualitatively predict the observed trends that the smallest fraction of metals is retained in stars for lower mass galaxies with an increase at higher masses, and that the fraction retained in the ISM is a factor of a few greater than in the stars. Quantitatively, the fraction of metals retained in stars in simulations from \citet{Muratov2017} and \cite{Piacitelli2025} are in reasonable agreement with measurements of the stars from GLOW galaxies, in contrast to the simulations from \cite{Christensen2018} where the retention fractions are a factor of a few higher than we measure in real galaxies. Similarly, the metal retention predicted in the ISM from all three simulations is higher than observed, with the largest disagreement coming from \citet{Muratov2015}. \citet{Christensen2018} also noted the discrepancy between the metal retention in the ISM and attributed part of the differences to the larger physical scale of what is considering `ISM' in simulations vs.\ the more modest region that can be sampled observationally. These authors note that a more careful comparison of spatial scales in the simulations (i.e., a more limiting definition of the ISM) reduces, but does not remove, the differences. This suggests that the physics governing the full baryon cycle and feedback processes in the simulated galaxies are not accurately capturing the detailed chemical evolution of galaxies and our empirical results provide new constraints to help aid improvements. Finally, we note the higher predicted metal retention fractions in the ISM changes the balance of where metals reside and may have implications for the amount of metals predicted to be in the CGM or lost to the IGM. It is worth noting that all three sets of simulations reproduce the observed MZ scaling relations for gas and for stars, highlighting the limited utility of scaling relations as discriminators for simulations.

\subsection{Interpretation Using a Regulator-Type Model of Galaxy Evolution}
\label{sec:regulatory_model}
Here, we use a simple regulator-type framework of galaxy evolution of \citet{Kravtsov.Manwadkar.2022}, which models galaxy processes in a simplified but realistic fashion, to help interpret the results of observations reported in this paper and the results of the simulations discussed above. The specific goal is to infer what our measured retained oxygen fractions imply about key galaxy formation processes, such as wind properties and preventative feedback, while considering whether the model that reproduces these fractions can also match observational estimates of wind mass loading factors in dwarf galaxies and the stellar mass-halo mass relation (SMHM) indicated by observed luminosity functions of Milky Way satellites and nearby field dwarf galaxies. 

The model evolves stellar and gas masses, metallicities, galaxy sizes, star formation rates, and other properties of the galaxies using a host halo evolution track computed using either an analytic approximation or a mass assembly history of a specific halo extracted from a cosmological simulation. The evolution is computed by solving a system of coupled differential equations connecting these properties using physically-motivated parameterizations of key properties. For example, the mass of heavy elements in stars, $M_{\rm Z,\star}$, is governed by a simple equation $\dot{M}_{Z,\star}=Z_{\rm g}(1-R)\dot{\mathcal{M}}_\star$, where $Z_{\rm g}=M_{Z,\rm g}/M_{\rm g}$ is the metallicity of the interstellar gas defined as ratio of the masses of heavy elements and gas, $R$ is the total mass fraction returned to the ISM by a single-age stellar population, and $\dot{\mathcal{M}}_\star$ is the star formation rate.

The mass of metals in the ISM, on the other hand, is governed by the equation for the rate of change of mass of heavy elements, which is due to accretion of gas from the intergalactic medium (IGM), $\dot{M}_{\rm Z,in}$, the production of new heavy elements by stars, and by the removal of heavy elements from the ISM when it is converted into stars or carried away in outflows (i.e., winds, which we denote using a subscript `w'): 
\begin{eqnarray}
&\dot{M}_{Z,\rm g} = \dot{M}_{\rm Z,in} + y_Z\dot{\mathcal{M}}_\star - Z_{\rm g}(1-R)\,\dot{\mathcal{M}}_\star - \eta_{\rm w} Z_{\rm w}\dot{\mathcal{M}}_\star\nonumber\\
&=Z_{\rm IGM}\dot{M}_{\rm g,in} +\left[y_Z - \left(1-R+\zeta_{\rm w}\eta_{\rm w}\right)\,Z_{\rm g}\right]\dot{\mathcal{M}}_\star,\label{eq:mzgevo}
\label{eq:mzgmodel}
\end{eqnarray}
where $Z_{\rm IGM}$ is the characteristic metallicity of heavy elements accreted by the galaxy, $y_Z$ is the yield of metals produced by massive stars and deposited into the ISM via the instantaneous recycling approximation, 
$Z_{\rm w}$ is characteristic metallicity of the gas carried out in winds, and $\zeta_{\rm w}=Z_{\rm w}/Z_{\rm g}$ is the wind metallicity enhancement factor accounting for differences in the metallicity of the wind and the ISM. 

Note that this model assumes that gas accretes onto galaxy directly from the IGM without any additional mediation by the CGM. In this model, IGM accretion and properties of winds have a large and direct impact on the star formation history and other properties of galaxies. This is different from recent models discussed in \citet[][]{Carr2023, Pandya2023, Voit2024}, in which the CGM mediates gas accretion and thereby greatly reduces the sensitivity of the galaxy properties to the wind mass loading factor.  However, galaxy formation simulations show that in halos of total mass $\lesssim 10^{11}\, M_\odot$ that host dwarf galaxies such as those in the GLOW mass range, gas is expected to be accreted via cold flows on an approximately free-fall time scale and no stable CGM is expected to form \citep{Keres2005,Keres2009,Dekel.Birnboim.2006,Brooks2009}. In the regime relevant for the galaxies in the GLOW sample the assumption in our model is therefore justified. 

\begin{figure}
\includegraphics[width=0.45\textwidth]{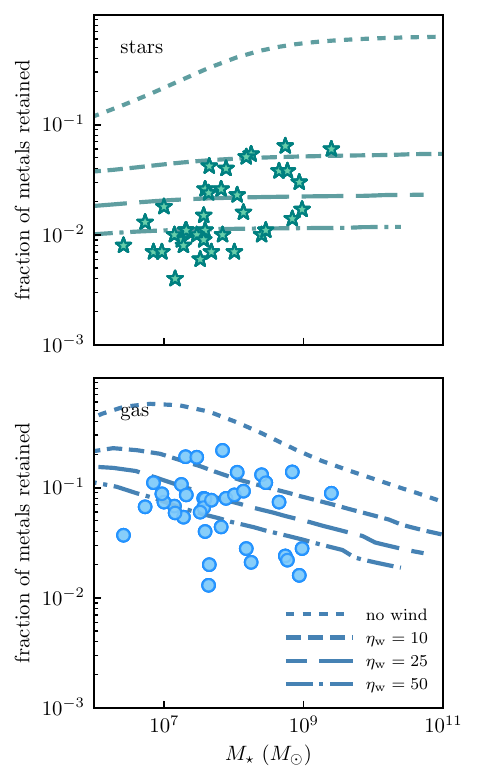}
\caption{The fraction of retained oxygen in the modelled low-mass galaxies (lines) and in the GLOW sample (points). The top and bottom panels show the retained fraction in stars and ISM gas, respectively. The models assume a constant wind mass-loading factor $\eta_{\rm w}$ with the values from 0 (no wind) to 50, as indicated in the legend, and the wind metallicity equal to that of the ISM gas, $\zeta_{\rm w}=1$. The assumption that mass-loading factors are constant do not reproduce the observed trend that the oxygen retained in stars is dependent on a galaxy's stellar mass.}
\label{fig:fret_eta_const}
\end{figure}

This galaxy formation model reproduces a wide range of properties of observed dwarf galaxies, including metallicity-luminosity, size-luminosity relation, luminosity function and radial distribution of the dwarf satellites of the Milky Way satellites down to faintest luminosities \citep[][]{Kravtsov.Manwadkar.2022,Manwadkar.Kravtsov.2022,Kravtsov.2024}. It is thus sufficiently realistic for the purpose of interpreting the results of observations and simulations.

For simplicity, we model the evolution of galaxy properties using the mean mass assembly history for halos of a given final mass $M_{\rm h}$ using an accurate approximation to the mean halo mass accretion rate for halos of different mass and at different epochs in simulations \citep[see Appendix in][for details and tests]{Kravtsov.Belokurov.2024}. We use this approximation to generate mass assembly histories for a grid of halo masses and calculate the galaxy evolution for each halo mass in the grid. This generates predicted properties of galaxies as a function of halo mass. 

In what follows, we use the model described in \citet{Manwadkar.Kravtsov.2022} but with an additional modeling of dust evolution following the model of \citet{Feldmann.2015}. We fix the values of $R=0.433$ and $y_Z={\rm P}=0.01$ adopted in the analyses of observations above. We keep all other parameters at the values presented in \citet{Manwadkar.Kravtsov.2022}, except for the wind mass loading factor, $\eta_{\rm w}$, the metallicity of the outflow relative to that of the ISM, $\zeta_{\rm w}$, and the parameter controlling the rate of accretion of the IGM gas, $\epsilon_{\rm pr}$, which allows us to model effects of IGM preheating \citep[see \S 2.2 in][for details]{Kravtsov.Manwadkar.2022}. 

We start with the assumption that the wind metallicity is equal to the ISM metallicity (i.e, $\zeta_{\rm w} = 1$). Such an assumption is reasonable in the case of strong winds (i.e., large mass loading factors $\eta_{\rm w}$) when a large fraction of the ISM is removed by the wind. 
Figure~\ref{fig:fret_eta_const} shows the fraction of retained metals in stars (top panel) and gas (bottom panel) for the models where the mass loading factor $\eta_{\rm w}=\rm constant$ and $\zeta_{\rm w}=1$\footnote{Note that for the case of $\eta_{\rm w}=\rm constant$, retained metal fraction is degenerate to the changes of $\eta_{\rm w}$ and $\zeta_{\rm w}$ as can be seen in Equation~\ref{eq:mzgmodel}.} compared to the values obtained for the observed galaxies in the GLOW sample. As expected, the retained metal fraction is sensitive to the value of the wind mass loading factor, $\eta_{\rm w}$, and stronger winds are needed to expel a larger total mass of metals. However, none of the $\eta_{\rm w}=\rm const$ models reproduce the observed trend of increasing the retained metal fraction in stars with increasing stellar mass, although $\eta_{\rm w}=25$ model roughly matches the characteristic value of the retained metal fraction in both stars and gas. To reproduce the trend of the retained metal fraction with stellar mass, $\eta_{\rm w}$ and/or $\zeta_{\rm w}$ must depend on $M_\star$. 

\begin{figure}
\includegraphics[width=0.45\textwidth]{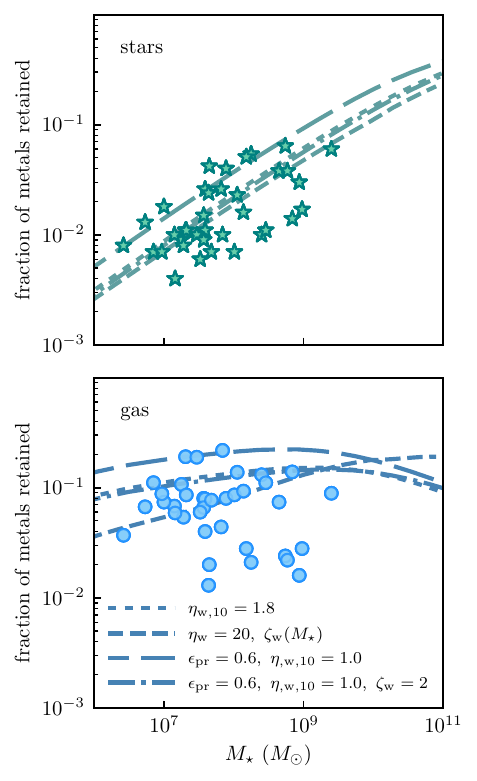}
\caption{The fraction of retained oxygen in model dwarf galaxies (lines) and in the GLOW sample (points). The top and bottom panels show the retained fraction in stars and ISM gas, respectively. Three of the four shown models models assume $\zeta_{\rm w}=1$ (i.e., wind has the same metallicity as the ISM), but the  wind mass loading factor depending on galaxy stellar mass as $\eta_{\rm w}=\eta_{\rm w,10}(M_{\star}/10^{10}\, M_\odot)^{-0.45}$, where $\eta_{\rm w,10}$ is the normalization with the values indicated in the legend. The fourth model assumes constant mass loading functor $\eta_{\rm w}=20$ and the wind metallicity relative to the ISM metallicity scaling as $\zeta_{\rm w}=(M_{\star}/10^{8}\, M_\odot)^{-0.45}$. These models illustrate that the form of the observed trend can be reproduced by different model assumptions. Three of the shown models also reproduce the characteristic values of the retained metal fraction of the GLOW galaxies. }
\label{fig:fret_etaw_ms}
\end{figure}

Figure~\ref{fig:fret_etaw_ms} shows results for four such models with a mass-dependent component: three models have the mass loading factor dependent on stellar mass as a power law ($\eta_{\rm w}=\eta_{\rm w,10}\times M_{\rm star,10}^{-0.45}$, where $M_{\rm star,10}=M_\star/10^{10}\, M_\odot$) and one has the the wind metallicity enhancement dependent on stellar mass as a power law. Such a power law form for the mass loading factor is consistent with the expectation of energy-driven wind models \citep[see, e.g.,][]{Kravtsov.Manwadkar.2022} and results of the simulations \citep{Christensen2016, Muratov2017}. In all the models, suppression of the IGM gas accretion rate ($\dot{M}_{\rm IGM}$) is controlled by the parameter $\epsilon_{\rm pr}$ as
$\dot{M}_{\rm IGM}\propto\epsilon_{\rm pr}f_{\rm b}\dot{M}_{\rm h}$, where $f_{\rm b}=\Omega_{\rm b}/\Omega_{\rm m}$ is the universal baryon fraction and $\dot{M}_{\rm h}$ is the total halo mass accretion rate. In two of the models shown in Figure~\ref{fig:fret_etaw_ms}, $\epsilon_{\rm pr}=1$, while in the other two accretion is somewhat suppressed with $\epsilon_{\rm pr}=0.6$ (i.e., accretion to the galaxies is 60\% of the universal baryon fraction). We further discuss the impact of this parameter below.

In the first model, $\eta_{\rm w,10}=1.8$, which is motivated by the fact that this value and the mass-dependent model were shown to reproduce the luminosity function of Milky Way dwarf satellites and a wide range of other dwarf galaxy properties. Figure~\ref{fig:fret_etaw_ms} shows that the model $\eta_{\rm w,10}=1.8$ reproduces the observed trends of the retained metal fraction in stars and gas as a function of stellar mass. However, the numerical values of the mass-loading factors in this model range from 114 to 5 as \mstar\ increases from $10^6$ \msun\ to $10^9$ \msun, which is considerably higher than observational estimates of 0.02 to 20 for actively star-forming dwarf galaxies over this mass range in the nearby universe. Specifically, a recent study reported mass-loading factors of $\sim0.02$, with considerable scatter, for low-mass starbursting galaxies \citep{Marasco2023}. Mass loading factors of $\sim 0.1-3$ were found for field dwarfs identified in the SAGA survey \citep{KadoFong2025}. Values of 0.2 to 7 were reported in a sample of actively star-forming galaxies with \mstar\ between $10^7-10^9$ \msun\ \citep{McQuinn2019}. The mass loading factors in this study were found to be more closely correlated with the concentration of star formation than with the gravitational potential. Finally, somewhat higher values of $10-20$ have been reported in low-mass galaxies with stronger starbursts \citep{Heckman2015, Chisholm2017}. 

Thus, in Figure~\ref{fig:fret_etaw_ms} we explore two models with lower mass loading factors using $\eta_{\rm w,10}=1.0$. In the first of these models, we use the same value of $\zeta_{\rm w}=1$ as above. As the mass-loading factors are lower, fewer metals are expelled, and the model therefore shows higher metal retention fractions which are a poor match to the measurements from the GLOW galaxies. However, the effects of $\eta_{\rm w}$ and relative metallicity of the wind to the ISM, $\zeta_{\rm w}$, on the retained metal fraction are degenerate in the model. Thus, the observational estimates of the retained oxygen fraction can be reconciled in models with the lower values of $\eta_{\rm w}$ if $\zeta_{\rm w}>1$. In Figure~\ref{fig:fret_etaw_ms} we show such a model setting $\zeta_{\rm w}=2$, which brings the metal retention fractions into better agreement with empirical measurements. The metallicity of a wind can be greater than that of the ISM if newly synthesized heavy elements produced by young massive stars are lost with hot gas via an open channel created as the superbubbles driven by stellar winds and supernovae explosions break out of the ISM (see Section~\ref{sec:compare_mzr}). In this case, metals can be driven out of the galaxy before they have a chance to mix with the rest of the gas, thereby enhancing metallicity of the material being lost relative to the metallicity of the ISM. Values of $\zeta_{\rm w}\approx 0.5-2$ are indeed predicted in the FIRE-2 simulations of dwarf galaxies \citep{Muratov2017}.

To further highlight the complexity of the parameter space, we show a final model in Figure~\ref{fig:fret_etaw_ms} with a fixed $\eta_{\rm w}=20$ and a mass-dependent $\zeta_{\rm w}=(M_\star/10^8\,M_\odot)^{-0.45}$ where $\zeta_{\rm w}$ values vary from 2.5 to $0.4$ over the stellar mass range of the GLOW sample. This model produces similar metal retained fractions to the model with $\eta_{\rm w}=1.8M_{\rm star,10}^{-0.45}$, although there are some differences due to the fact that massive galaxies in this model retain smaller amount of gas and this affects their star formation rate. 

Clearly, as shown in Figure~\ref{fig:fret_etaw_ms}, there is not enough information to discern between these models if considering only metal retention fractions as a function of stellar mass alone, but the degeneracy between the models can be broken if we also consider their implications for the stellar mass function (or luminosity function) of galaxies. To reproduce the observed stellar mass (or luminosity) function of the Milky Way satellites or dwarf galaxies in the field, galaxies in the model should occupy halos of the correct halo mass, $M_{\rm h}$ and the model must predict a correct $M_\star-M_{\rm h}$ relation. 

\begin{figure}
\includegraphics[width=0.45\textwidth]{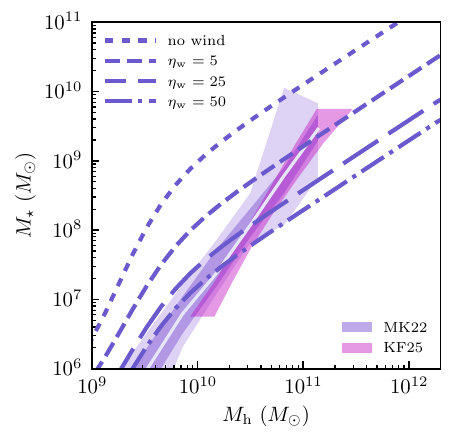}
\includegraphics[width=0.45\textwidth]{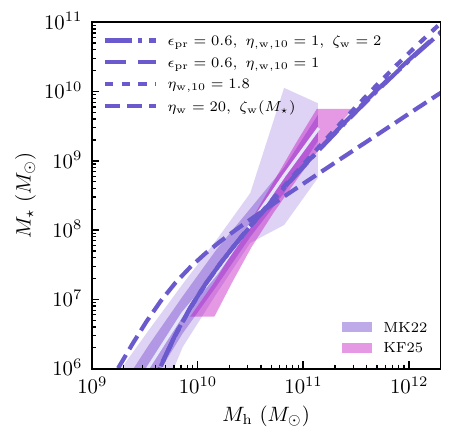}
\caption{Stellar mass--halo mass relations for the models shown  Figures~\ref{fig:fret_eta_const} and \ref{fig:fret_etaw_ms} (blue lines) compared with the relations inferred for the observed dwarf satellites of the Milky Way \citep[blue shaded regions showing $1\sigma$ and $2\sigma$ scatter, see][]{Manwadkar.Kravtsov.2022} and for nearby field dwarf galaxies \citep[magenta shaded region showing $2\sigma$ uncertainty][]{KadoFong2025}. The top panel shows that $\eta_{\rm w}=\rm const$ are inconsistent with the form of the $M_\star-M_{\rm h}$ relation inferred for observed dwarf galaxies. The bottom panel shows four models in which either wind mass loading factor, $\eta_{\rm w}$, or wind metallicity relative to the ISM, $\eta_z$, depend on stellar mass. Three of the four shown models models assume $\eta_{\rm z}=1$ (i.e., wind has the same metallicity as the ISM), but the  wind mass loading factor depending on galaxy stellar mass as $\eta_{\rm w}=\eta_{\rm w,10}(M_{\star}/10^{10}\, M_\odot)^{-0.45}$, where $\eta_{\rm w,10}$ is the normalization with the values indicated in the legend. The fourth model, assumes constant mass loading factor $\eta_{\rm w}=20$ and but the wind metallicity relative to the ISM metallicity scaling as $\zeta_{\rm w}=(M_{\star}/10^{8}\, M_\odot)^{-0.45}$. In two of the models the accretion of baryons is suppressed $\epsilon_{\rm pr}=0.6$ relative to the fiducial case of $\epsilon_{\rm pr}=1$ ($\dot{M}_{\rm g}\propto \epsilon_{\rm pr}f_{\rm b}\dot{M}_h$).}
\label{fig:msmh_eta_const}
\end{figure}

In Figure~\ref{fig:msmh_eta_const}, we show the $M_\star-M_{\rm h}$ relations that results from the models considered in Figures~\ref{fig:fret_eta_const} and \ref{fig:fret_etaw_ms}, overplotted on the relation that reproduces the observed luminosity function of the dwarf MW satellites \citep[see][]{Manwadkar.Kravtsov.2022} and the $M_\star-M_{\rm h}$ relation inferred for the nearby observed field dwarf galaxies by \citet{KadoFong2025}. The top panel shows the $\eta_{\rm w}=\rm constant$ models (from Figure~\ref{fig:fret_eta_const}) and makes it clear that such models do not reproduce the relations inferred for dwarf galaxies in observations for any value of $\eta_{\rm w}$. The bottom panel shows the models with stellar mass dependent $\eta_{\rm w}$, as well as the model with a higher constant $\eta_{\rm w}$ value and stellar mass dependent $\zeta_{\rm w}$ (from Figure~\ref{fig:fret_etaw_ms}).
We can see that $\eta_{\rm w}(M_\star)$ models reproduce the shape of the $M_\star-M_{\rm h}$ relation inferred from observations, while the $\eta_{\rm w,10}=1.8$ model (the same model that matches many other properties of dwarf galaxies) also matches the amplitude of the inferred relation reasonably well. The model with $\eta_{\rm w}=20$ and $\zeta_{\rm w}=(M_\star/10^8\,M_\odot)^{-0.45}$, on the other hand, predicts an incorrect form of the $M_\star-M_{\rm h}$ relation and overestimates $M_\star$ for the faintest galaxies and significantly underestimates $M_\star$ of larger dwarf galaxies, such as Large Magellanic Cloud. This is simply because $\zeta_{\rm w}$ does not affect star formation significantly, while $\eta_{\rm w}$ does and the model with $\eta_{\rm w}=20$ produces incorrect $M_\star-M_{\rm h}$ relation just as the other such models in the top panel of this figure. Thus, the stellar mass dependence of the ratio of the wind metallicity relative to the metallicity of the ISM, $\zeta_{\rm w}(M_\star)$, cannot be the explanation for the observed trends in the retained metal fraction in stars. This trend and the form of the $M_\star-M_{\rm h}$, however, can be explained by the stellar mass dependence of mass loading factor, $\eta_{\rm w}\propto M_\star^{-0.45}$. 

The bottom panel of Figure~\ref{fig:msmh_eta_const} also shows the two models with mass loading factors that are mass dependent ($\eta_{\rm w}(M_\star)$; these two models have accretion of the IGM gas suppressed by $\epsilon_{\rm pr}=0.6$ while normalization of $\eta_{\rm w}(M_\star)$ is lowered from 1.8 to $\eta_{\rm w,10}=1.0$. The suppression of gas accretion can be due to preventative feedback, whereby IGM is heated by the previous generation of winds of the galaxy and its neighbors to temperatures above the virial temperature of the halos hosting dwarf galaxies \citep[e.g.,][]{Bennett2025}. One of the models assumes $\zeta_{\rm w}=1$, while the other assumes $\zeta_{\rm w}=2$. Figure~\ref{fig:msmh_eta_const} shows that these models produce identical $M_\star-M_{\rm h}$ relations and that these relations are similar to the relation in the model with $\epsilon_{\rm pr}=1$ and $\eta_{\rm w,10}=1.8$. In other words, the lower mass loading factors are able to reproduce similar stellar mass as higher mass loading factors since there is overall less gas accreted from the IGM to fuel star formation. Figure~\ref{fig:fret_etaw_ms} also shows that the model with $\epsilon_{\rm pr}=0.6$, $\eta_{\rm w,10}=1.0$, and $\zeta_{\rm w}=2$ is a better match to the average values of the observed retained metal fractions of GLOW galaxies while avoiding mass-loading factors that are significantly higher than observed. 

To summarize, the results presented in this section demonstrate that the retained metal fractions in dwarf galaxies are set by the properties of the gas outflows, in particular their mass loading factor, the metallicity of the outflows relative to the metallicity of the ISM, and the accretion rate of the IGM gas. This illustrates why the task of reproducing the observed retained metal fractions and their trend with stellar mass in simulations is challenging, as it requires correct modeling of all these processes. For example, if preheating of the IGM is important, as indicated by results discussed above, zoom-in simulations of individual dwarf galaxies may not account for it if they do not model or resolve feedback and heating of the IGM from the nearby galaxies. 

The observed trend of the retained metal fraction in stars with stellar mass likely reflects the increase of the wind mass loading factor with decreasing stellar mass. In particular, we showed that the scaling $\eta_{\rm w}\propto M_\star^{-0.45}$, expected in the energy-driven winds, is consistent both with the retained metal fraction trend exhibited by the GLOW sample and in the $M_\star-M_{\rm h}$ relation inferred from the luminosity function of Milky Way satellites and nearby field dwarf galaxies, if normalization of the relation at $M_\star=10^{10}\, M_\odot$ is set appropriately. This normalization can be reconciled with the observational estimates of $\eta_{\rm w}$ for nearby galaxies in the stellar mass range of $M_\star\sim 10^7-10^9\,M_\odot$ in the model where the IGM gas accretion is moderately suppressed due to preventative feedback and wind mass loading factor normalization is reduced by a factor of two. 

Note that to achieve the suppression of the IGM gas by a factor of two, the IGM should be heated to temperatures corresponding to the virial temperature of the halo. For dwarf galaxies in the mass range of $M_\star\sim 10^7-10^9\, M_\odot$, galaxies should be hosted in halos of $M_{\rm 200}\sim 10^{10}-10^{11}\, M_\odot$ at $z=0$ (see Figure~\ref{fig:msmh_eta_const}), which corresponds to virial temperatures of $T_{200}\sim 0.5-2 \times 10^5$ K. However, a large fraction of stars formed at higher $z$ and so accretion suppression should have been occurring at higher $z$ as well. For a given halo mass, the required IGM temperature will increase as $T_{\rm 200}\propto (1+z)$ at larger redshifts. The preventative feedback via preheating preferred by the considerations discussed above thus implies significant temperature of the IGM in the regions that have evolved into our local cosmic neighborhood. 

\section{Conclusions}\label{sec:conclusions}

In this work, we have measured the production, distribution, and retention of oxygen in a sample of 37 gas-rich, star-forming low-mass galaxies in the local universe (D$<$6 Mpc) using multi-wavelength data obtained primarily from the astronomical archives and presented in Paper I (McQuinn et al.\ 2026). Here, we summarize our main conclusions:

\begin{itemize}
\item We find that most low-mass ($10^{6.5}$ \msun\ $<$ \mstar\ $< 10^{9.5}$ \msun) galaxies retain only a small fraction of the oxygen produce by stellar nucleosynthesis with only a modest decrease for the lowest mass galaxies in the sample. The percent of oxygen retained ranges from 4$-$25\% with mean of 10.7\%. The majority of the oxygen is retained in the gas, with a much smaller reservoir found in the stars and stellar remnants and in the dust. Based on comparison with existing studies, the oxygen fraction increases quite quickly (by a factor of $3-4$) for galaxies with \mstar\ $\gtsimeq10^{10}$ \msun, and the majority of oxygen is retained in the stars rather than the ISM \citep{Peeples2014, Telford2019} (see Section~\ref{sec:mrf}; Figure~\ref{fig:orf_parsec}).
\item The combined constraints that the galaxies have a low retention of oxygen but high gas content can be understood within the framework in which metals are preferentially expelled via supernovae, in hot-phase winds moderately enhanced in metallicity compared to the ISM, while the ISM gas mass is regulated by mass-loss in warm winds with characteristic wind mass loading factors averaged over cosmic time of $\eta_{\rm w}\sim 1-10$ in the stellar mass range of the GLOW sample. 
\item The low-mass galaxies at larger distances ($>1$ Mpc) from a more massive neighbor (\mstar $>10^9$ \msun) consistently retain a low percentage of their oxygen ($<$15\%). For galaxies found in closer proximity to their nearest large  galaxy, there is increased scatter in the percent of oxygen retained, indicating environment impacts the amount of oxygen retained, recycled, or even accreted to galaxies. We also note that that the galaxies that are both in closer proximity {\em and} that reside in an overall higher density environment (measured by the tidal index $\Theta_5$) tend to have higher retention fractions. While based on a small sample of galaxies, if this tentative trend is confirmed, this would suggest there is a greater degree of oxygen that is recycled back to these galaxies (likely due to higher ambient pressure in the local environment) and/or greater accretion of metals that were previously expelled by other galaxies in the local environment.
\item Contrary to naive expectations, the oxygen retained in the galaxies does not correlate with the distribution of galaxies in the MZ relation, nor with galaxy stellar mass or gas-phase oxygen abundance. The amount of oxygen retained in low-mass galaxies showed a modest increase in galaxies with higher stellar mass, greater gas content, and deeper gravitational potential traced by the full-width half maximum of the 21 cm line (see Section~\ref{sec:mrf}; Figure~\ref{fig:orf_parsec}).
\item Previous studies of the metal content in the CGM of local low-mass galaxies, including some of the galaxies included in the GLOW sample, estimate only a small fraction ($1-7$ \%) of the oxygen produced by stars resides in the hot halo; extrapolating to include cool/warm gas phases based on ionization correction factors, the oxygen budget in the CGM may be 5$\times$ that, or 5$-$35\%. This would suggest a significant fraction of the oxygen has been lost from the gravitational potential of the galaxies, but given the current level of uncertainties, it is still an open question whether the majority of oxygen produced by stars in the low-mass galaxies has been lost to the IGM and is a primary source of IGM `pollution'. 
\item We compared our empirical constraints to results from three separate cosmological simulations and found that while all simulations reproduce the changing distribution of metals retained in stars vs.\ the ISM in galaxies as a function of galaxy mass, the amount of metals retained in the simulated galaxies were higher than what is measured by observations. The simulations also predict a notable increase in retention fraction at higher masses, which is not seen in the combined sample of galaxies considered here and in the literature. As the simulations all reproduce well the observed MZ relations at these galaxy masses, the distinct differences in the retention of metals on an individual galaxy level highlight the importance of studying the detailed phenomenon and the limitations in scaling relations. 
\item Using a simple regulator-type galaxy evolution model, we explored different mass-loading factors, wind metallicity enhancements, and accretion rate of IGM gas to galaxies to simultaneously reproduce empirical constraints on mass-loading factors, metal retention fractions, and the stellar mass-halo mass relation. We find that the data are well matched by a model in which the wind mass loading factor depends on the stellar mass of the galaxy as $\eta_{\rm w}\propto M_\star^{-0.45}$, the metallicity of expelled material is twice that of the ISM, and the amount of material accreted onto these low-mass systems is suppressed by 40\% (i.e., the baryon content acquired is less than the universal baryon fraction). We show that this model reproduces both the retained metal fractions in low-mass galaxies and the stellar mass--halo mass relation indicated by the observed luminosity function of Milky Way satellites and nearby field dwarf galaxies. 
\end{itemize}

\section{Acknowledgments}
KBWM thanks Molly Peeples for providing the results from \citet{Peeples2014} for inclusion in Figure~\ref{fig:orf_parsec} and Daniel Piacitelli for providing the oxygen retention fractions from \citet{Piacitelli2025} for inclusion in Figure~\ref{fig:orf_models}. The authors would like to the thank the referee for helpful comments that improved this manuscrupt. This work was supported by the National Science Foundation under grant numbers 1806926 and 1940800 (PI McQuinn). Additional support came from programs HST-GO-16144 (PI McQuinn) and HST-GO-15227 (PI Burchett), provided by NASA through a grant from the Space Telescope Science Institute, which is operated by the Associations of Universities for Research in Astronomy, Incorporated, under NASA contract NAS5- 26555. 
O.\ G.\ T.\ acknowledges support from grant HST-AR-16155 from the Space Telescope Science Institute and from a Carnegie-Princeton Fellowship through Princeton University and the Carnegie Observatories.
This research has made use of NASA’s Astrophysics Data System Bibliographic Services and the NASA/IPAC Extragalactic Database (NED), which is operated by the Jet Propulsion Laboratory, California Institute of Technology, under contract with the National Aeronautics and Space Administration. The Flatiron Institute is funded by the Simons Foundation.

\facility{Hubble Space Telescope, Very Large Array,  Australian Telescope Compact Array, Spitzer Space Telescope}

\appendix{}
\section{Analysis using the MIST Stellar Libary}\label{sec:appendix}
Here, we repeat the \ORF\ analysis using the SFHs derived from the same HST data but using the MIST stellar libraries instead of PARSEC (see Appendix in Paper~I).  In Figures~\ref{fig:orf_mist_properties} and \ref{fig:orf_mist} we present the results for direct comparison to Figures~\ref{fig:orf_properties} and \ref{fig:orf_parsec}. Overall the results are in very good agreement with those derived using the PARSEC library. There are subtle differences for individual galaxies which indicate the level of systematic uncertainties in the analysis; however, given that the stars retain the smallest fraction of metals of all galactic components in low-mass systems, the impact on the \ORF\ calculations is minimal. 

\begin{figure}
\begin{center}
\includegraphics[width=0.415\textwidth]
{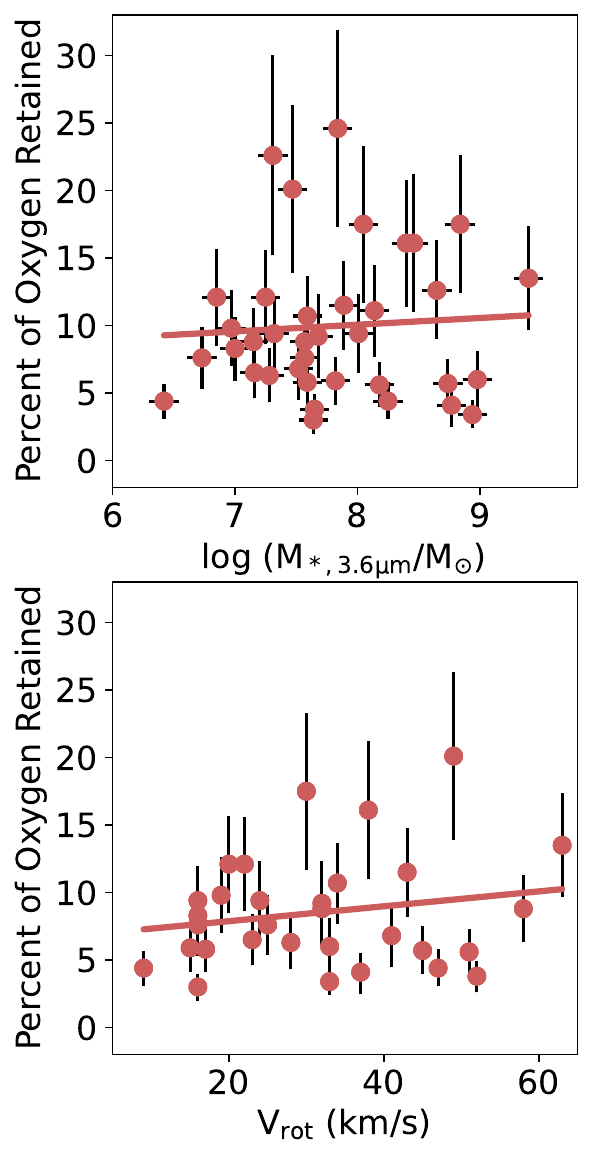}
\includegraphics[width=0.48\textwidth]{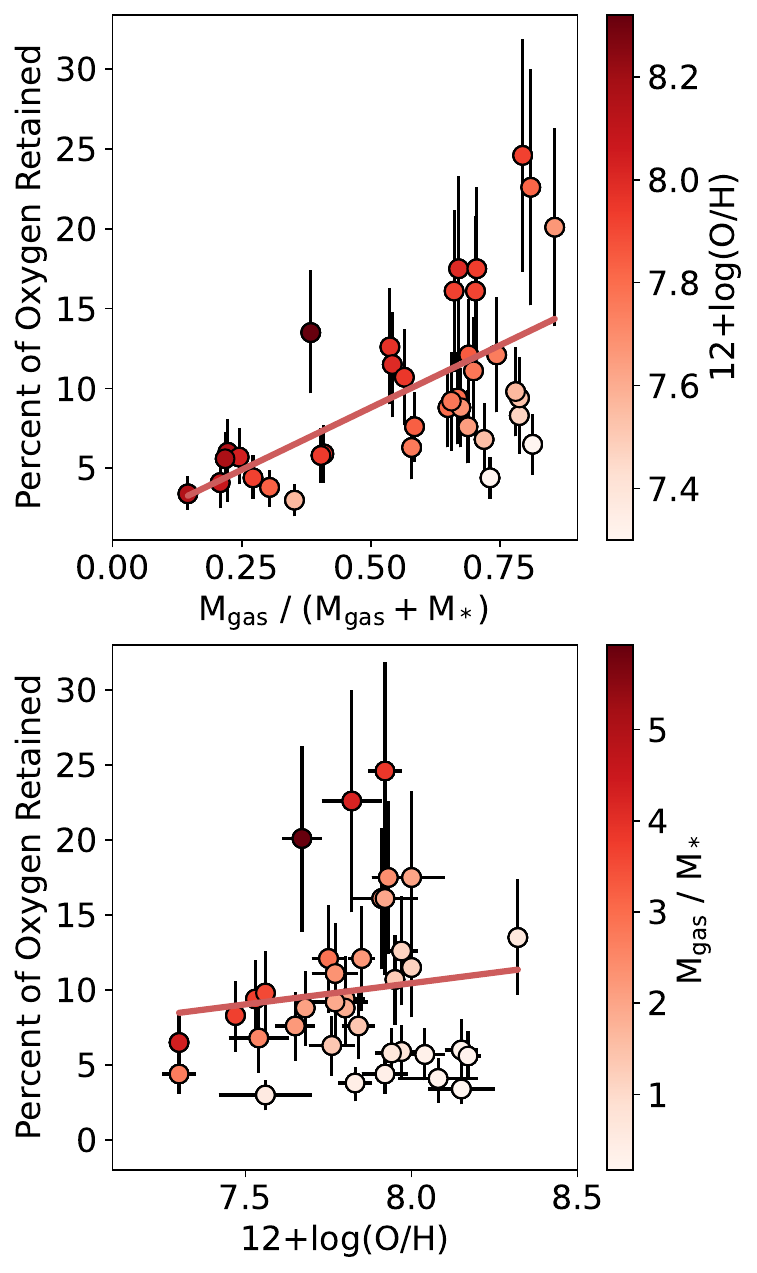}
\end{center}
\caption{Same as Figure~\ref{fig:orf_properties} but based on SFHs and AMRs derived using the MIST stellar library.}
\label{fig:orf_mist_properties}
\end{figure}

\begin{figure*}
\includegraphics[width=0.9\textwidth]{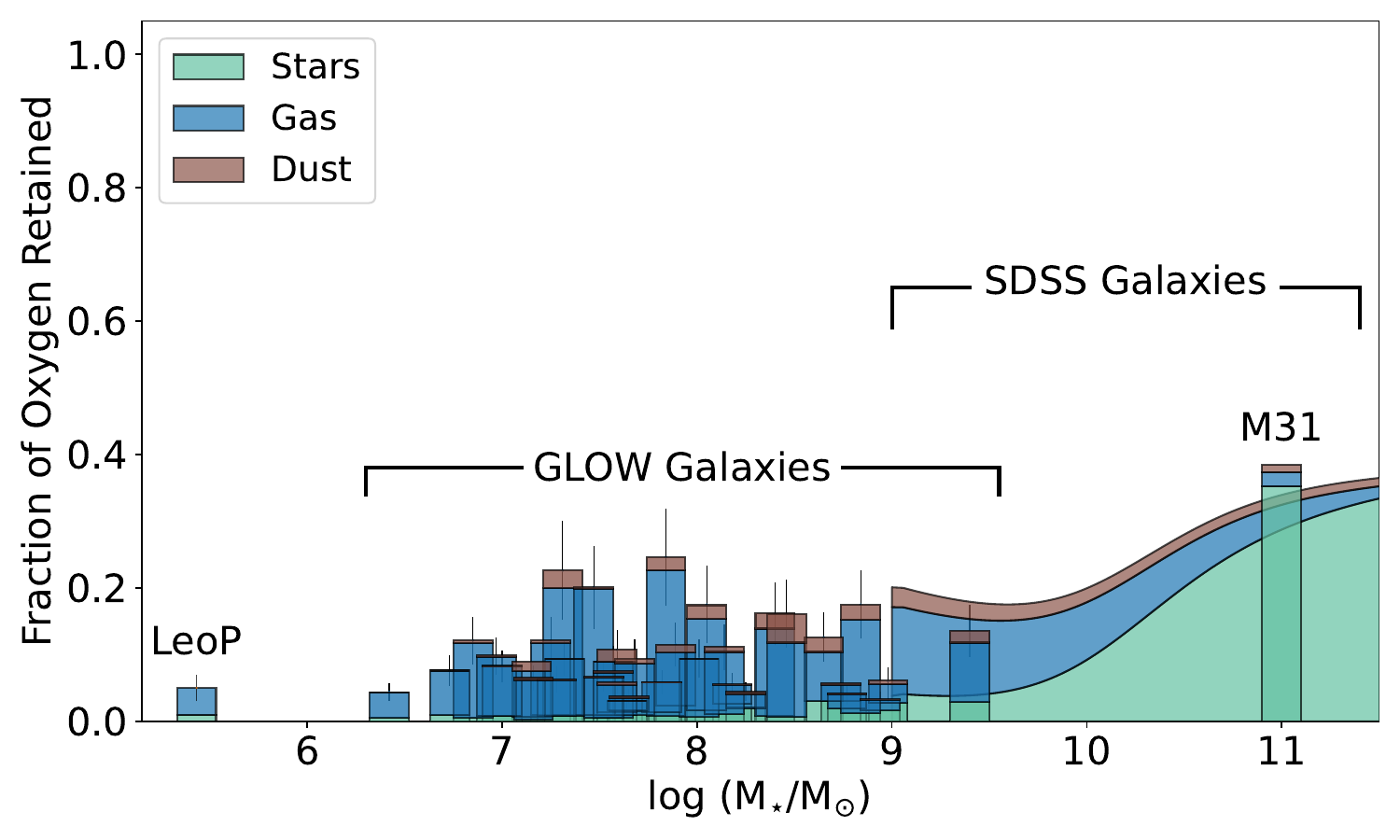}
\caption{Same as Figure~\ref{fig:orf_parsec} but based on SFHs and AMRs derived using the MIST stellar library.}
\label{fig:orf_mist}
\end{figure*}

\renewcommand\bibname{{References}}
\bibliography{ms_mrf.bib}
\end{document}